\documentclass[fleqn,usenatbib]{mnras}

\usepackage{newtxtext,newtxmath}

\usepackage[T1]{fontenc}

\DeclareRobustCommand{\VAN}[3]{#2}
\let\VANthebibliography\thebibliography
\def\thebibliography{\DeclareRobustCommand{\VAN}[3]{##3}\VANthebibliography}

\usepackage{graphicx}	
\usepackage{amsmath}	
\let\oldAA\AA
\renewcommand{\AA}{\text{\normalfont\oldAA}}

\newcommand{\oi}{[\ion{O}{I}]}

\newcommand{\oiii}{[\ion{O}{III}]}

\newcommand{\nii}{[\ion{N}{II}]}

\newcommand{\sii}{[\ion{S}{II}]}

\newcommand{\hii}{\ion{H}{II}}

\newcommand\met{$12 + log(O/H)$}
\newcommand\ha{H$\alpha$}
\newcommand\hb{H$\beta$}

\newcommand\hcm{{\sc HII-CHI-Mistry}}

\title[]{A MUSE view of the multiple interacting system HCG\,31}

\author[D. A. G\'omez-Espinoza et al.]{
Diego A. G\'omez-Espinoza,$^{1}$\thanks{E-mail: diego.gomez@userena.cl}
S. Torres-Flores,$^{1}$, V. Firpo$^{2}$, Philippe Amram$^{3}$, Benoit Epinat$^{3,5}$, \newauthor Thierry Contini$^{4}$ and Claudia Mendes de Oliveira$^6$
\\
$^{1}$ Departamento de Astronomía, Universidad de La Serena, Av. Juan Cisternas 1200 Norte, La Serena, Chile\\
$^{2}$ Gemini Observatory/NSF's NOIRLab, Casilla 603, La Serena, Chile\\
$^{3}$ Aix Marseille Univ, CNRS, CNES, LAM, Marseille, France.\\
$^{4}$ Institut de Recherche en Astrophysique et Planétologie (IRAP), Université de Toulouse, CNRS, UPS, CNES, 31400 Toulouse, France.\\
$^{5}$ Canada-France-Hawaii Telescope, 65-1238 Mamalahoa Highway, Kamuela, HI 96743, USA\\
$^{6}$ Departamento de Astronomia, Instituto de Astronomia, Geofísica e Ciências Atmosféricas da USP, Cidade Universitaria, CEP: 05508-900, São Paulo, SP, Brazil.\\
}

\date{Accepted 2023 April 4. Received 2023 April 4; in original form 2022 August 8}

\pubyear{2023}

\begin{document}
\label{firstpage}
\pagerange{\pageref{firstpage}--\pageref{lastpage}}
\maketitle

\begin{abstract}
We present, for the first time, spatially resolved spectroscopy for the entire Hickson Compact Group 31 obtained with the MUSE instrument at the VLT,and an in-depth analysis of this compact group. To obtain a complete understanding of the system, we derived radial velocity and dispersion velocity maps, maps of the ionization mechanism of the system, chemical abundances and their distribution over the whole system, star formation rates and ages of the different star-forming regions, and the spatial distribution of the Wolf-Rayet stellar population. We also reconstructed the star formation history of the galaxies HCG 31 A, C, B and F, measured the emission-line fluxes, and performed a stellar population synthesis. Our main findings are: (i) that there is clearly disturbed kinematics due to the merger event that the system is experiencing; (ii) that the ionization is produced exclusively via star formation except for the nucleus of the galaxy HCG 31 A, where there is a small contribution of shocks; (iii) that there is low oxygen abundance distributed homogeneously through the system; (iv) that there is a prominent population of carbon Wolf-Rayet stars in the central zone of the group; and (v) that there are clear evidences of the tidal origin of the galaxies HCG 31 E, HCG 31 H, and HCG 31 F because they show quite high oxygen abundances for their stellar mass. All these findings are clear evidence that HCG 31 is currently in an early merging phase and manifesting a starburst in its central region.

\end{abstract}




\section{Introduction}
\label{introduction}

The processes of formation and transformation of galaxies are crucial for understanding their evolution. 
According to the hierarchical model of galaxy formation, the interactions between small galaxies at high redshift are the foundations for the formation of current galaxies \citep{toomre&toomre72}, therefore, a detailed study of local interacting/merging galaxies can provide important information regarding different phenomena that were common in the distant universe.
A great advantage of studying galaxy mergers and interactions in the local universe is the high spatial resolution that we have due to the proximity of these objects. Although various studies have been done comparing the properties of galaxies in the local and distant universe (\citealt{epinat+10}, \citealt{Perez-montero+21}, \citealt{Izotov+21}), it should be noted that there are clear differences between studying galaxy mergers in the local universe and studying at high redshift. For example, the mass of the gas has evolved with redshift and also the star formation rates (SFR) (\citealt{Mannucci+10}, \citealt{Behroozi+13},\citealt{Madau+14}, \citealt{Amorin+15},  \citealt{Boyett+22}). Thus, the comparison of physical processes and phenomena is more relevant than the comparison of the absolute values. Galaxy mergers are unique laboratories for understanding the transformation of galaxies due to gravitational effects. However, to fully understand these phenomena, it is imperative to combine observational results with simulations and models.

There are many types of mergers between galaxies. Depending on the type of collision, the galaxies will be affected in different ways. If one galaxy is much less massive than the other ($< 1:4$ of the mass according to \citealt{Kaviraj14}), the merger is called a \textit{minor merger}. This type of merger does not observationally affect the most massive galaxy. If the two galaxies have similar mass (or at least one has $> 1:3$ of the mass of the other according to \citealt{Kaviraj13}) a \textit{major merger} is said to have occurred in which both galaxies are affected in their morphology. There are some features that indicate that a major merger is occurring, such as tidal tails, destruction of the disks, formation of tidal dwarf galaxies (TDGs) and flattening of the metallicity gradient, among others (\citealt{toomre&toomre72}, \citealt{Duc&mirabel98}, \citealt{Kewley+10}, \citealt{rich+12},\citealt{de_mello+12}, \citealt{torres-flores+14}, \citealt{Mora+19}, \citealt{Torres-flores+20}).

During the past few years, several authors have studied interacting galaxies by analyzing their physical properties (\citealt{zaragoza-cardiel+10}), as well as from the point of view of their kinematic properties (\citealt{Plana+03}, \citealt{amram+07}, \citealt{torres-flores+10}). These efforts have allowed us to understand the role of galaxy interactions in their evolution, however, several questions remain open, with no clear answers. For example, because of the interaction between galaxies an increase in their SFR was expected.  Various authors have proposed different scenarios to explain this increase. \cite{Xu+10} studied a sample of pairs of spiral-elliptical galaxies and found that most spiral galaxies did not show an increase in their SFRs. On the other hand, \cite{Patton+11} studied a sample of galaxy pairs and found that they did increase their SFRs, which was expected due to the interaction processes between galaxies. Additionally, \cite{Ellison+11} found that galaxies associated in pairs had a higher fraction of active galactic nuclei (AGN) compared to isolated galaxies. \cite{rupke+10} found that interacting galaxies exhibited more flattened metallicity gradients than those observed in isolated galaxies, a phenomenon that could be produced by the interaction between galaxies.

There are other processes in galaxy interactions that result in the formation of new galaxies. \cite{Duc&mirabel98} studied the interacting system NGC 5291, where they found star-forming objects in the intergalactic medium. The authors claimed that this suggested the existence of newly formed tidal dwarf galaxies in NGC 5291. These objects should be free of dark matter and should have high metallicities (considering their masses). In addition to the TDGs detected in NGC 5291, other authors have detected and studied TDG candidates in different interacting systems (\citealt{Weilbacher+03}, \citealt{de_mello+12}).

As shown before, interacting and merging systems are ideal laboratories for understanding the different physical and kinematical processes that take place in our universe. Various observational techniques have been used over the years to investigate these systems, such as optical imaging, longslit spectroscopy, Fabry-Perot data, HI data cubes, and recently, 3D spectroscopy. The integral field spectroscopy technique (IFS) is one of the most powerful approaches for studying an interacting/merging system in detail thanks to its usually wide spectral range, especially if the field of view is large enough to cover all members, as with a compact group of galaxies, where gravitational encounters are quite common.

In this work we developed a deep spectroscopic study of the Hickson Compact Group 31, which is a complex interacting system of dwarf galaxies located in the nearby universe \citep{rubin+90}. Several authors (\citealt{iglesias-paramo&vilchez97}, \citealt{lopez-sanchez+04}, \citealt{mendesdol+06}, \citealt{amram+07}, \citealt{gallagher+10}, \citealt{alfaro-cuello+15}, \citealt{Torres-flores+15}) have studied this object but none of them have combined a well-adapted spatial coverage with a broad spectral coverage, which is the case for the MUSE data presented in this work. Thus, this paper provides, for the first time, a complete view of this group, based on MUSE/VLT data of the main galaxies of the system and of the southern tidal tail. Therefore, the analyses presented in this work can be very valuable in the study of galaxy transformation and evolution in dense environments. The paper is organized as follows: in section \ref{hcg31} we present the system; in section \ref{data} we present the data; in section \ref{analysis} we present the analysis used to derive the emission-line intensities, kinematics, extinction, star formation rates (SFR), oxygen abundance, equivalent width (EW), ages, and ionization mechanism; and, in section \ref{results} we present our results. The discussion and summary are presented in sections \ref{discussion} and \ref{summary}, respectively.  

\section{The Hickson Compact Group 31}
\label{hcg31}

Our object of study consists of a specific group of galaxies, the Hickson Compact Group 31 (RA [Deg] = 75.409572, DEC [Deg] = -4.257011). This group lies at a distance of 59.38 $\pm$ 4.16 Mpc. That distance was measured assuming a redshift of $z = 0.01347 \pm 0.00002$ \citep{wong+06} and a value of the Hubble constant of $H_{0} = 67.8 km \ sec^{-1} \ Mpc^{-1}$ \citep{Riess+16}. 

This group consists of many low-mass galaxies with low-metallicities \citep{mendesdol+06}, and all of them are in an interacting event \citep{rubin+90}. This particular configuration makes HCG\,31 an ideal laboratory to study galacticy interactions and evolution.

Due to its configuration, HCG\,31 is a very well-studied object and many authors have contributed information about this system, (e.g \citealt{rubin+90}, \citealt{iglesias-paramo&vilchez97}, \citealt{lopez-sanchez+04}, \citealt{amram+07}, \citealt{alfaro-cuello+15}, \citealt{Torres-flores+15}, among others). \cite{hickson82} classified it as a compact group of galaxies. In this study the authors detected four members in this group, HCG\,31 A, B, C and D. \cite{rubin+90} detected four new members, HCG\,31 E, F, G and Q, and concluded that member D was not part of the group due to its higher redshift. More recently, \cite{mendesdol+06} detected another member of the system, HCG\,31 R.
Currently, we accept the evidence that HCG\,31 consists of nine galaxies: A, B, C, E, F, G, H, Q and R. Figure \ref{fovs} presents an optical image of the group with eight of the members labeled (member R is not seen due to its low brightness).

The entire group is embedded in a common envelope of neutral hydrogen, which has a total HI mass of $2.1 \times 10^{10} \  M_{\odot}$ \citep{williams+91}. These authors also found that the HI distribution peaked at the overlap region between HCG\,31 A and HCG\,31 C.

\cite{lopez-sanchez+04} performed a complete study of this system. They developed a deep analysis of the physical properties by using optical imaging, near-infrared (NIR) imaging, and optical medium-resolution long-slit spectroscopy. An interesting result obtained by these authors was the detection of a Wolf-Rayet bump in the spectrum of the galaxy HCG\,31 C, which indicates the presence of Wolf-Rayet stars with very young ages (< 4 Myrs). This bump can be identified with the blend of the spectral lines He$_{II} \ \lambda  4686 \ \AA$, C$_{III}$/C$_{IV} \  \lambda  4650 \ \AA$ and N$_{III} \ \lambda \lambda  4634,4640 \ \AA$. Moreover, the most important emission in the bump arises from the helium lines. The presence of such stars is strong evidence of a young starburst in this system. In addition, these authors found that all the members of the group displayed low oxygen abundance, spanning a range of \met $\sim$ 8.03-8.37. 

\cite{mendesdol+06} derived the luminosity-metallicity relation (LZR) for HCG\,31 using K$_{s}$-band magnitudes, which mainly traced stellar emission (at this redshift). The LZR obtained by the authors suggested that galaxies C, G and B followed that relationship. Members H, R, E and F showed higher metallicities for their luminosities, which led the authors to suggest that these members could be tidal debris or TDG candidates, because this type of object retains the metallicity of its parent galaxy (\citealt{Weilbacher+03}).

A deep study of the central region of HCG\,31 was done by \cite{alfaro-cuello+15}. The authors used IFS observations taken with GMOS/Gemini, centered in the overlap region between galaxies A and C, and found a high SFR density in the central region. Using the same data \cite{Torres-flores+15} detected a flat gradient in the oxygen abundance map linking galaxies A and C. This gradient suggested gas mixing in the interface between galaxies A and C \citep{Torres-flores+15}. \cite{alfaro-cuello+15} also detected a super star cluster (SSC) near the nucleus of galaxy HCG\,31 C. These authors speculated that this SSC was currently triggering star formation in their surroundings.

As shown above, HCG\,31 offers an ideal place to seek for answers to different open questions in the field of interacting/merging galaxies.

\section{Observation and data reduction}
\label{data}

\begin{figure}
    \centering
    \includegraphics[scale=0.62]{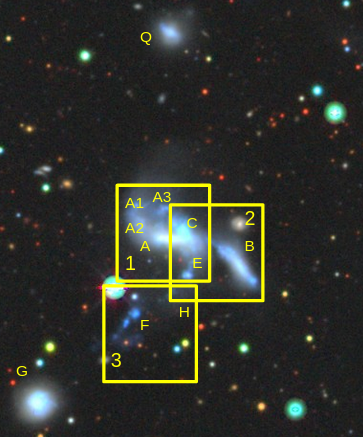}
    \caption[Test]{Optical image of HCG\,31. The yellow squares show the location of the IFUs taken by MUSE. Image taken from DECAls legacy survey.\footnotemark}
    \label{fovs}
    \vspace{0cm}
\end{figure}

The Hickson Compact Group 31 was observed during the nights of February 18-20 of 2014, as part of the science verification strategy. To obtain a larger spatial coverage, three different FoVs were observed. In Figure \ref{fovs} we represent all the FoVs observed over an optical image of HCG 31. The data was acquired in the Wide Field Mode (WFM), where each field covers a field-of-view of 1x1 arcmin$^2$, with a spatial sampling of 0.2\,arcsec  and a spectral range of 4750\,-\,9350\,\AA\, for a 1.25\,\AA\ step. The mean seeing during the observations was $\sim$\,0.9\,arcsec, with a mean airmass of 1.15. 
Considering the distance to the system (59.38 Mpc) our mean seeing implies a spatial resolution of $\sim$\,252\,pc, where each spaxel covers $\sim$\,56\,pc, and each FoV cover an area of $\sim$\,16.8\,x\,16.8\,kpc.

\footnotetext{\url{http://www.legacysurvey.org/}}

The data was reduced using the standard ESO pipelines for MUSE data reduction.

\section{Analysis}
\label{analysis}

In this section we present the analyses carried out to determine the different physical parameters and properties of the members of HCG\,31. We focus on the methods used to obtain extinction, SFR, the mechanism of ionization, and the measurement of line-emission fluxes, stellar population synthesis (SPS) and kinematics.

\subsection{Emission-line measurement}

The emission lines have been measured using two codes: {\sc FADO}\footnote{\url{http://spectralsynthesis.org/index.html}} (fitting analysis using differential evolution optimization, \citealt{Gomes+17}) and {\sc ifscube}\footnote{\url{https://ifscube.readthedocs.io/en/latest/}} \citep{Ruschel-dutra&oliveira20}. The first is designed to perform spectral population synthesis analysis to derive different physical parameters from a galaxy spectrum, including emission-line fluxes. The second code is a {\sc Python}\footnote{Python Software Foundation. Python Language Reference, version 2.7. Available at \url{http://www.python.org}} package of spectral analysis routines which fit Gaussian profiles to emission lines. We used {\sc ifscube} to obtain the radial velocity and velocity dispersion maps because {\sc FADO}s not optimized for such calculations. Due to the spectral resolution of the MUSE data (average R $\sim$3000), we fitted a single Gaussian profile to the observed line instead of performing a multi component analysis.

\begin{figure*}
    \includegraphics[scale=0.5]{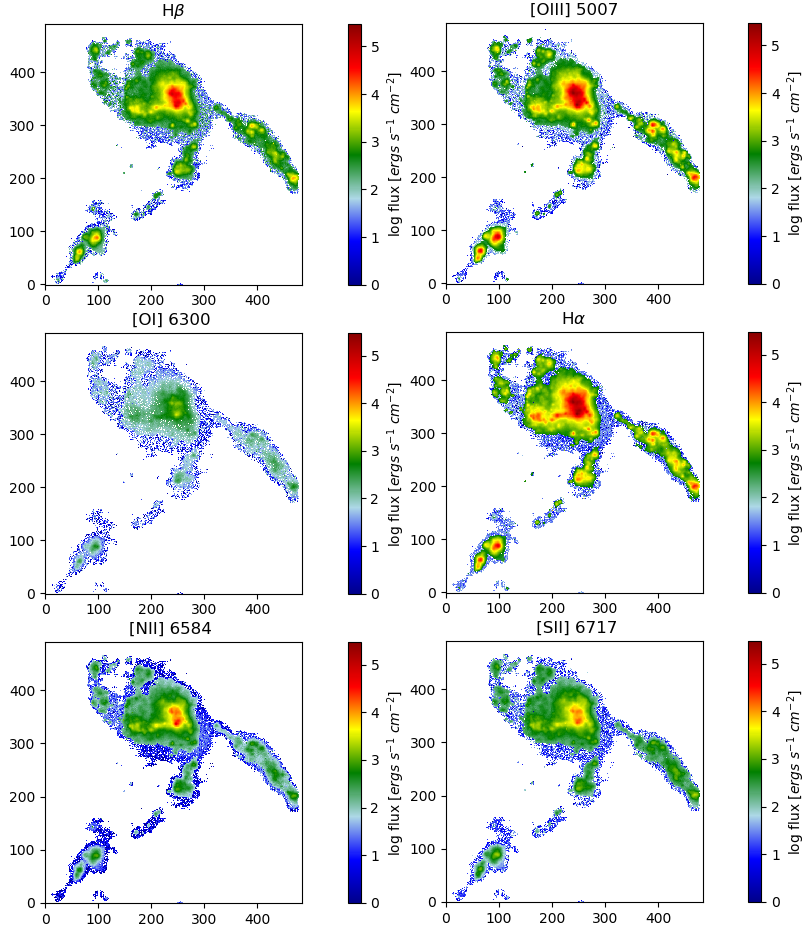}
    \caption{Emission-line flux maps of HCG31 obtained with {\sc FADO}. The lines are H$\beta$, \oiii\,$\lambda$\,5007\,\AA, \oi\,$\lambda$\,6300\,\AA, H$\alpha$, \nii\,$\lambda$\,6584\,\AA, and \sii\,$\lambda$\,6717\,\AA. The maps are in pixel coordinates and the scale color represents the flux. All maps are normalized to the maximum \ha \  flux.  }
    \label{linemaps}
    \vspace{0.5cm}
\end{figure*}

In Figure \ref{linemaps} we show the line-emission maps obtained for \hb, \oiii $\lambda$ 5007 $\ \AA$ , \oi $\lambda$ 6300 $\ \AA$, \ha, \nii $\lambda$ 6584  $\ \AA$ and \sii $\lambda$ 6717 $\ \AA$, which were derived from FADO analysis on the entire data cube. We used FADO output to reconstruct the maps in the (x,y) plane. These maps correspond to the flux obtained from the Gaussian fit performed by FADO.

Maps were cleared of noise by including spaxels where the signal-to-noise ratio (SNR) was $>3$. To measure the SNR we divided the map of the emission line with the map of the same line but in the variance layer of the datacube. Emission line fluxes were corrected for galactic and internal extinction by using the extinction laws proposed by \cite{Fitzpatrick99} and \cite{Calzetti00}, respectively.



\subsection{Kinematics: Radial velocity and velocity dispersion }


Kinematic analysis was done by using {\sc ifscube} on the H$\alpha$ emission line. The uncertainties of radial velocities were calculated using the formulas of \cite{Lenz+92} and the fitting parameters were derived by using {\sc ifscube}.

The width of a Gaussian profile is usually determined by the full width at half maximum (FWHM) which, in this case, is measured in angstroms. The standard deviation is related to the FWHM through FWHM\,=\,2$\sqrt{2ln2}$ $\times$ $\sigma$ \,$\approx\,2.35$ $\times$ $\sigma$. 

Under the assumption that the H$\alpha$ emission line can be fitted by a Gaussian profile, its width ($\sigma$) represents the velocity dispersion of the atoms, providing information about its broadening mechanism. The intrinsic velocity dispersion of the ionized gas ($\sigma_{int}$) is defined by equation \ref{eq1}:


\begin{equation}
    \sigma_{int}^{2} = \sigma_{obs}^{2} - \sigma_{inst}^{2} - \sigma_{th}^{2}
    \label{eq1}
\end{equation}

where $\sigma_{obs}$ is the observed width, $\sigma_{inst}$ is the instrumental width and $\sigma_{th}$ is the thermal broadening which can be estimated by using $\sigma_{th}\,=\,\sqrt{k_{B}T / m_{a}}$, where $k_{B}$ is the Boltzmann constant, $T$ is the electronic temperature and $m_{a}$ corresponds to the mean atomic mass of the gas. In our case, we assumed a standard $T\,=\,10{^4}K$, which is a good approximation for the temperature calculated by \cite{lopez-sanchez+04} for HCG\,31 ($T\,\sim\,9400\,\pm\,600\,K$ for the central zone); therefore we used $\sigma_{th}\,=\,9.4\,km s^{-1}$. For the instrumental width we use the standard $\sigma_{inst}$ for MUSE, which has been taken from the literature (\citealt{bellocchi+19}, $\sigma_{inst}\,=\,50\,kms^{-1}$ for the \ha \ wavelength). The uncertainties of $\sigma_{obs}$ were calculated according to the formula given by \cite{Lenz+92} with the fitting parameters of {\sc ifscube}.

 The stellar kinematic analysis were not perfomed due to the instrinsic weak continuum of the spaxels, which makes the fit of the stellar absorption lines difficult.


\subsection{Oxygen abundances}

The metallicity distribution in spiral galaxies has been extensively studied during the last years (\citealt{vanzee+98} \citealt{Bresolin+12}, \citealt{sanchez+14}). Most of giant galaxies show a clear abundance gradient, with the center of the galaxy being more metallic than the outskirts. On the other hand, several observational studies have proven that interacting galaxies show flatter abundance gradient. Owing to the spectral range of our data it was not possible to derive the oxygen abundance using the direct method because we could not measure the auroral line \oiii\,$\lambda$\,4363\,\AA. In this case, we decide to use the strong-line methods as oxygen abundance indicator, given that these methods mainly use strong emission-lines which typically have quite high SNR. Specifically, we use the N2 and O3N2 empirical calibrators which requires only four intense emission-lines: H$\alpha$, H$\beta$, \oiii\,$\lambda$\,5007\,\AA\ and \nii\,$\lambda$\,6584\,\AA. 

We use the calibrations obtained by \cite{marino+13} for the N2 and O3N2 calibrators








We also use the code {\sc HII-CHI Mistry} \citep{Perez-montero14}
which is a collection of python scripts that analyze the intensities from several bright emission lines observed in
the optical (in our case \ha, \hb, \nii \ $\lambda 6584 \ \AA$ and \oiii \ $\lambda 5007 \ \AA$). This code estimates the oxygen abundance by using grids of photoionization models.

\subsection{Star formation rate determination}

The SFR is a key parameter in galaxy evolution and represents the amount of gas that is converted into stars per unit of time. Several authors have studied how the environment plays a role in the star formation rate of interacting/merging systems (\citealt{Mihos+96}, \citealt{Boselli+06},\citealt{Xu+10},\citealt{Teyssier+10},\citealt{Patton+11},\citealt{Pontzen+17}).



\cite{Kennicutt&evans12} reviewed the most used SFR calibrators to date, and they refined the equations, including for $H\alpha$. They propose, for $L(H\alpha)$, the following expression, assuming a Chabrier IMF, and a continuous SF process: 

\begin{equation}
    log SFR (M_{\odot} yr^{-1}) = log L(H\alpha) (erg s^{-1}) - 41.27
\end{equation}

On this paper we use H$\alpha$\ as tracer of SFRs. The uncertainties are calculated by propagating the flux errors given by {\sc FADO} and using an uncertainty on the distance of 4.16 Mpc (explained in section \ref{hcg31}).

\subsection{H alpha equivalent width and age determination}

The H$\alpha$ equivalent width EW($H\alpha$) gives us a good estimation of the ratio between ionizing photons of massive stars and continuum photons of the underlying stellar population and it is commonly used to date star formation events. One of the models used to estimate ages is given by {\sc STARBURST99} (\citealt{Leitherer90}), which is used in this work, given that 
it was previously used in the analysis of HCG\,31 (\citealt{lopez-sanchez+04}, \citealt{alfaro-cuello+15}). Therefore, we will be able to compare our findings with these previous studies.

{\sc STARBURST99} generates models for the evolution of different properties of a single stellar population, considering an instantaneous burst or a continuous star formation process. The predictions are available for five different stellar metallicities: ($2Z_{\odot}$, $Z_{\odot}$,$0.4Z_{\odot}$, $0.2Z_{\odot}$, and $0.005Z_{\odot}$), and for three different initial mass functions, covering ages from $10^{6}$ to $10^{9}$ years. On this work we used a model for a Z = 0.004 and Z = 0.008 metallicities, based on the observed oxygen abundance of the system. 


\subsection{Ionization mechanism: BPT diagnostic diagrams}

Distinguishing the different ionization mechanism in a galaxy is a very important and challenging topic in extragalactic astronomy. \cite{Baldwin81} used a database of extragalactic objects to classify them according to their ionization mechanism. The predominant ionization mechanism in an extragalactic object could be i) photo-ionization by OB stars, ii) a power-law continuum source, and iii) shock wave heating \citep{Baldwin81}. In this context, diagnostic diagrams based on emission line ratios, for instance $I([O_{III}] 5007 \AA)$/$I(H\beta)$ vs $I([N_{II}] 6584 \AA)$/$I(H\alpha)$, have been extensively used during the last decades to determine ionization mechanisms (hereafter, BPT diagrams).



The limits which divide the different ionization mechanisms is not fully clear. \cite{Kewley+01} studied the properties of starburst galaxies using the codes \textit{PEGASE} and \textit{STARBURST99}, which allowed them to derive an upper limit for the starburst region in the BPT diagrams. 


\cite{Kauffmann+03} studied a sample of 22623 AGNs at $0.02 < z < 0.30$ taken from the \textit{Sloan Digital Sky Survey} (SDSS) with the objective to separate star-forming galaxies and AGNs. \cite{Kauffmann+03} used the BPT diagrams and derived an empirical limit to the star-forming sequence. This limit lies below the line suggested by \cite{Kewley+01}, and the zone between both limits corresponds to the so-called composite region, which probably host a mixture between the different ionization mechanisms.

\subsection{Star formation History}

{\sc FADO} allows us to reconstruct the Star Formation History (SFH) of some object by fitting a combination of Single Stellar Population (SSP) to the spectrum of the galaxy, obtaining a population vector. This population is based on the \cite{Bruzual&Charlot03} stellar libraries. In order to calculate the ages and metallicities of the best-fitting vector {\sc FADO} substracts the nebular continuum from the observed spectra. The resulting stellar population is used to visualize the SFH of the galaxy by the luminosity at the normalization wavelenght of the different SSPs  (see \citealt{Gomes+17} for details). For HCG\,31, we use 5100 $\AA$ as the normalization wavelength.

\section{Results}
\label{results}

\subsection{Radial Velocity: The complex velocity field of HCG 31}
\label{radvel_results}

\begin{figure*}
    \centering
    \includegraphics[width=0.501\textwidth]{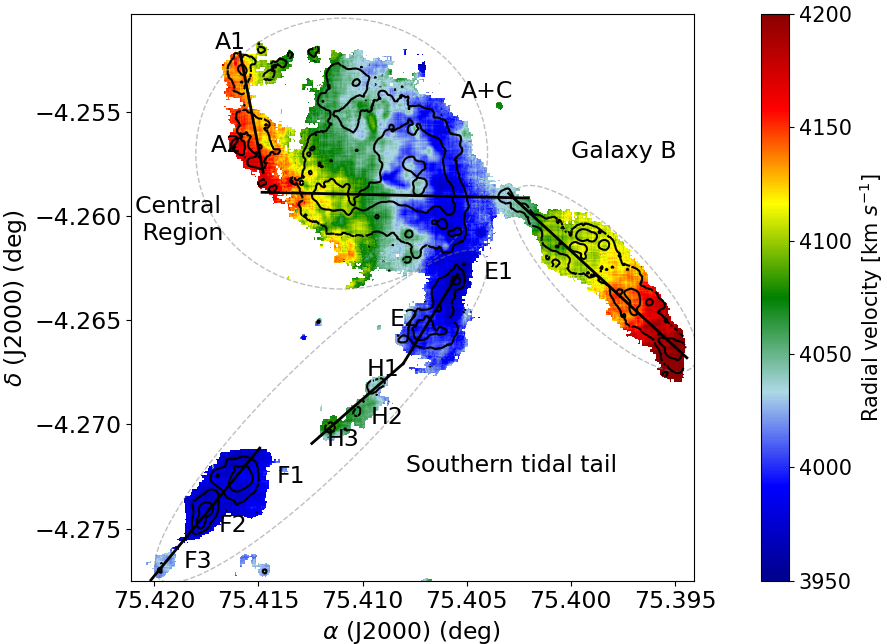} \includegraphics[width=0.487\textwidth]{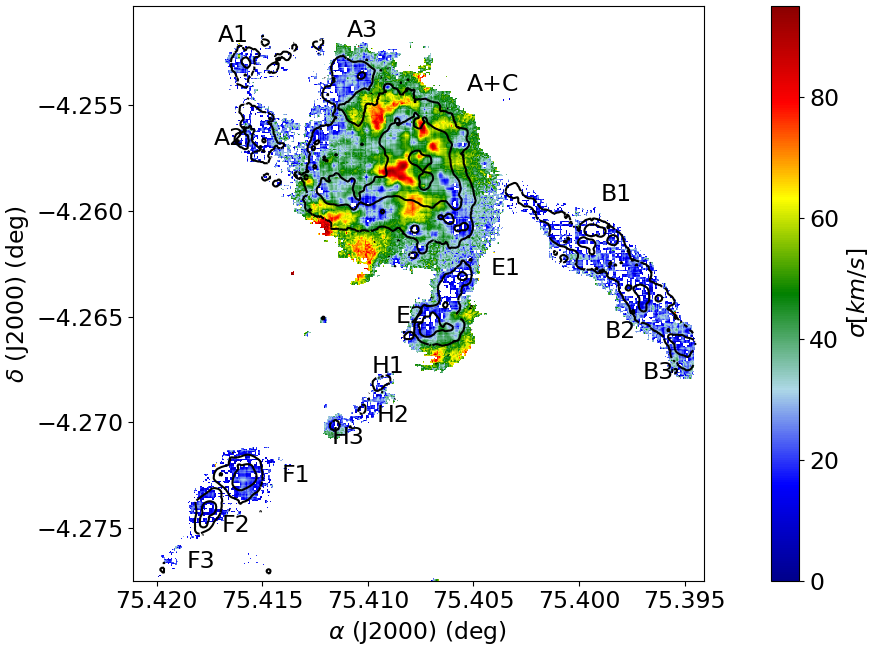} 
    \caption{Left panel: H$\alpha$ radial velocity derived by {\sc ifscube}. The black contours represent H$\alpha$ emission. The gray ellipses show the three main kinematic entities visually identified (see text). The black lines represent the different slits used to analyze the internal kinematics of each object. 
    Right panel: H$\alpha$ velocity dispersion map. The zones with higher velocity dispersion are in the center of the merger between A and C. Members B, E, and F show low velocity dispersion (< 30 km s$^{-1}$). The group spans a range of velocity dispersion of 10 km s$^{-1}$ < $\sigma$ < 95 km s$^{-1}$. However, our calculations probably overestimated the values (see text). In both panels north is up and east is on the left.}
    \label{map_vel}
\vspace{0.5cm}

\end{figure*}

In Figure~\ref{map_vel} (left panel) we display the radial velocity field of HCG\,31, derived from the H$\alpha$ emission line. The contours represent H$\alpha$ emission at the flux levels of 6.3$\times$10$^{-14}$ erg s$^{-1}$ arcsec$^{-2}$, 4.0$\times$10$^{-13}$ erg s$^{-1}$ arcsec$^{-2}$ and 1.0$\times$10$^{-11}$ erg s$^{-1}$ arcsec$^{-2}$. The velocity scale spans a range from 3950 km s$^{-1}$ to 4200 km s$^{-1}$. On this Figure, the main kinematic structures are labeled and the black lines represent the mock slits that we used to derive the position-velocity diagrams.

HCG\,31 shows a complex kinematics, and we could not assign a single rotating pattern for the whole system (\citealt{amram+07} hereafter A07). We identified three different kinematic entities: i) the central region (A+C), (ii) a western member called galaxy B, and (iii) the southern tidal tail composed of three sub-regions (E1, E2), (H1, H2, H3) and (F1, F2, F3). These three entities are labeled with gray ellipses in the left panel of Figure \ref{map_vel}.
The group covers a very narrow range in velocity space ($\sim$\,200\,km s$^{-1}$), which was also reported by other authors (\citealt{rubin+90}, \citealt{Richer+03}, \citealt{lopez-sanchez+04}, \citealt{amram+07}). This indicates that the system has a very low velocity dispersion, which is expected for compact groups composed of late-type (or gas-rich) galaxies \cite{hickson+88}.    

The central region shows a rotating pattern from East to West,
with a position angle (PA)\,$\sim 120^{\circ}$. A07 derived different kinematic parameters for this region. They estimated a mass of approximately $4.5 \times 10^{9} M_{\odot}$, PA\,=\,130\,$\pm$\,3$^{\circ}$, and an inclination of 52\,$\pm$\,5$^{\circ}$. The amplitude in radial velocity for this central region is about $\sim$ 150\,km s$^{-1}$. To analyze the velocity gradient of galaxy A, we simulated the PA used by \cite{verdes-montenegro+05} in the analysis of the HI map of HCG\,31 (Figure \ref{map_vel}, left panel, horizontal line in region A+C). The velocity gradient, shown in Figure~\ref{pseudoslits}, displays values similar to those presented by previous researchers.

{}

It should be noted that the slit passes through the zone where the H$\alpha$ double components are seen in the high-resolution Fabry-Perot data of this system (A07). Thus, the velocity gradient shown in figure~\ref{pseudoslits} corresponds to the average motion of the multiple emission line components, which cannot be resolved by MUSE. \cite{Richer+03} also showed Fabry-Perot data for HCG\,31, with a resolution of R\,$\sim$7500, concluding that A+C corresponded to a single kinematic entity. 

To confirm the kinematic nature of the central merger, we employed the MocKinG code\footnote{\url{https://gitlab.lam.fr/bepinat/MocKinG}} described in \cite{Mercier+22} which allows one to fit a rotating disk model to the velocity field with fixed projection parameters across the galaxy (position angle of major axis, center, inclination and systemic velocity). We used a Courteau rotation curve model \citep{courteau97} described by the following equation: 

\begin{equation}
    v(r) = v_c \frac{1+ (r_t/r)^\beta}{(1 + (r_t / r) ^ \gamma) ^ {(1 / \gamma)}} \text{~,}
    \label{eq:courteau}
\end{equation} 

 where $r$ is the radius, $r_t$ is the transition radius, $v_c$ is the asymptotic velocity, $\beta$ controls the steady rise at large radii, and $\gamma$ is related to the sharpness of the turnover after $r_t$. We use $\beta=0$ to reduce the number of free parameters as done in \citet{Gomez-Lopez+19}. We also fixed the center to a position inferred from the geometrical center of the continuum emission.

 MocKinG  uses a forward modeling approach that takes into account the limited spatial resolution of the data by weighting the velocity by the flux distribution within the point spread function and across each pixel  \citep[see Appendix of][for more details]{epinat+10}. It can use either a Levenberg-Markwardt algorithm \citep{Markwardt09} or the multimodal nest sampling algorithm Multinest \citep{Feroz+08, Buchner+14} to find the fit parameters. We used the latter solution that is much more robust to local minima.  

\begin{figure*} 
\centering
 \includegraphics[width=0.8\textwidth]{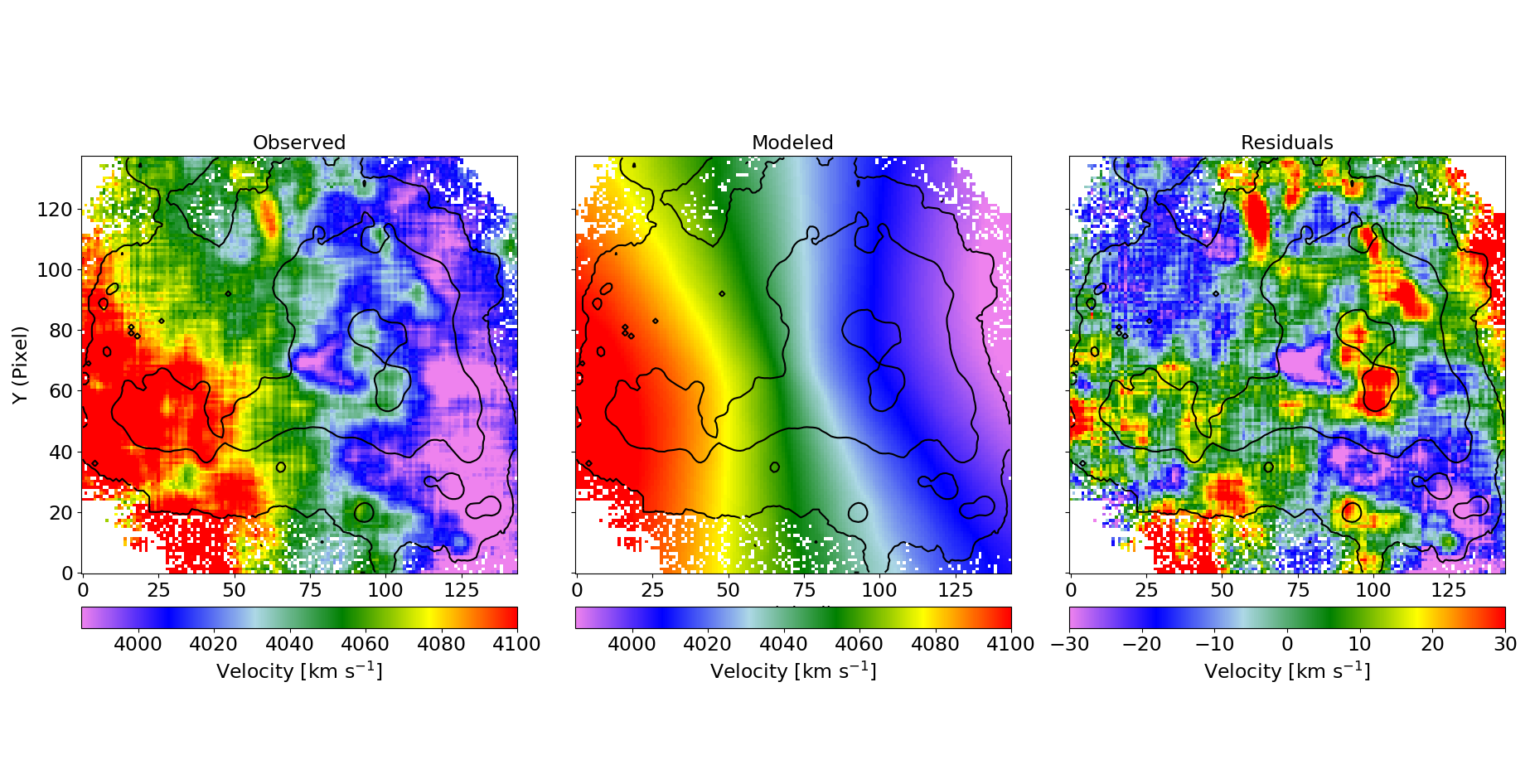}   \includegraphics[width=0.8\textwidth]{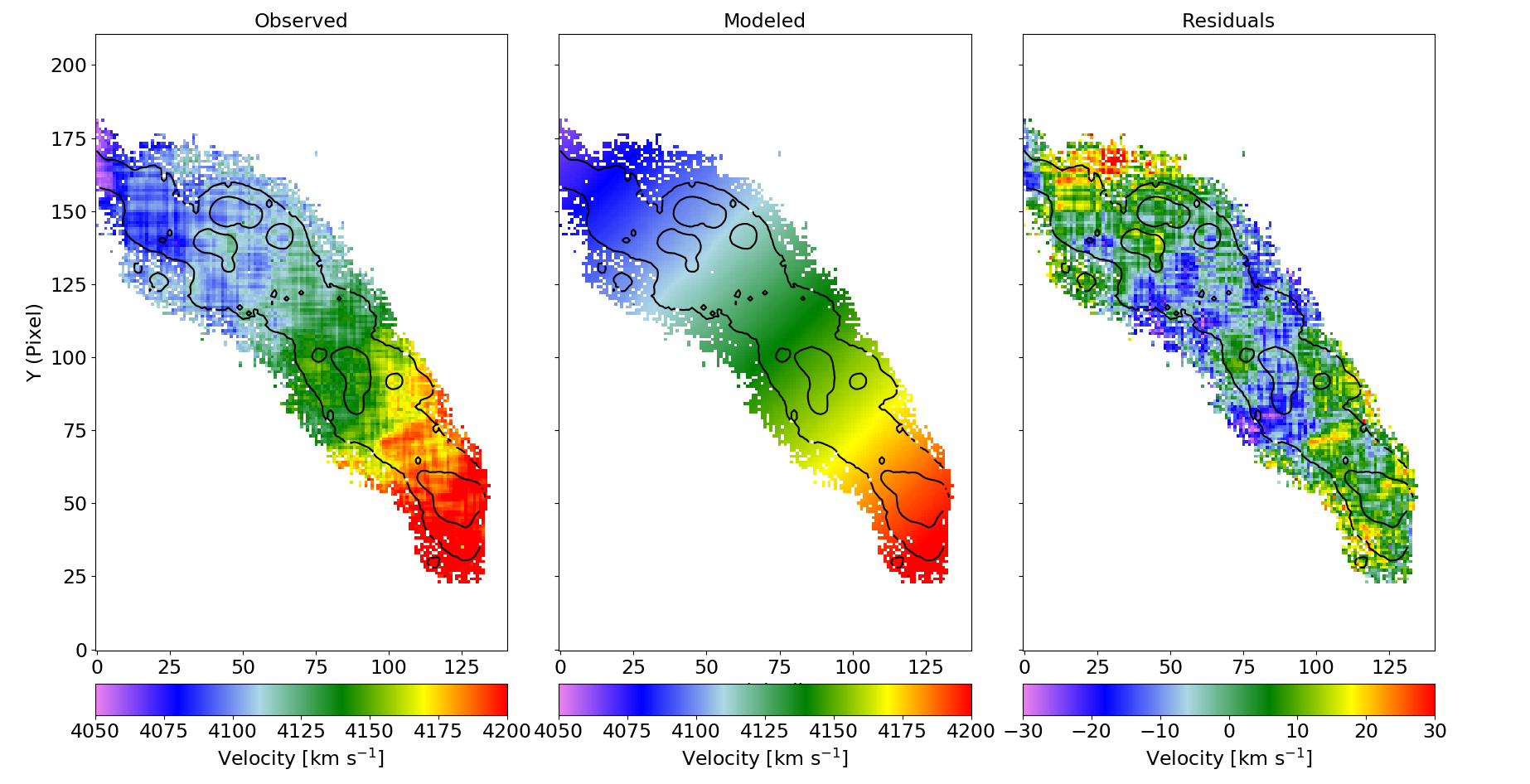}  
  \caption{ MocKinG results for the central merger (Top panel) and for Galaxy HCG 31 B (Bottom panel). Black contours represent \ha \ in emission.} 
 \label{barollo_results} 
\vspace{0.5cm}
\end{figure*}

\begin{figure*} 
\centering
 \includegraphics[width=0.45\textwidth]{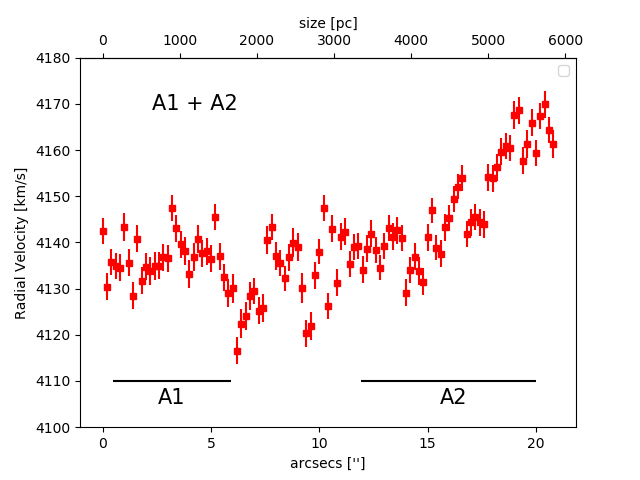}   \includegraphics[width=0.45\textwidth]{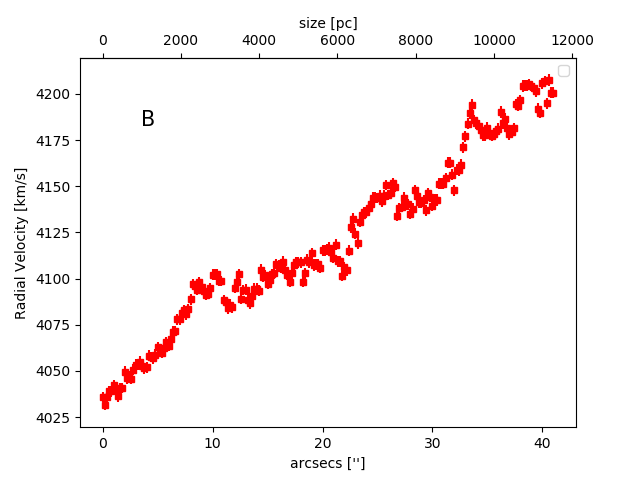}  
 \includegraphics[width=0.45\textwidth]{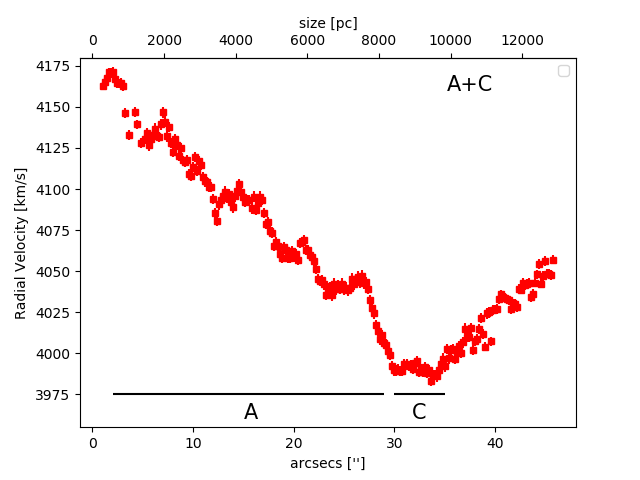} 
 \includegraphics[width=0.45\textwidth]{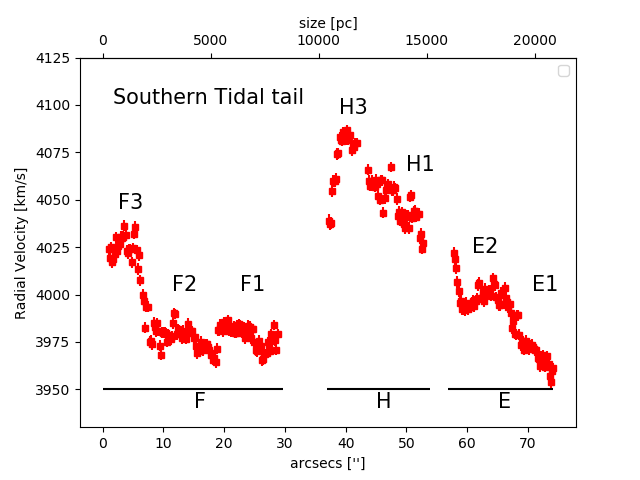}  \caption{Velocity gradients derived for the different pseudo slits represented in Figure~\ref{map_vel} (left panel). The gradients run from east to west.} 
 \label{pseudoslits} 
\vspace{0.5cm}
\end{figure*}

The resulting velocity field, model and residuals are shown in Fig. \ref{barollo_results} (top panel), where a clear pattern of rotation is observed.

The kinematics parameters obtained from the fit were: PA = 102\degr $\pm 1^{\circ}$ , $v_c=357 \pm 7$~km~s$^{-1}$, $r_t=14.4 \pm 0.3$\arcsec , $\gamma=0.69 \pm 0.1$, with an inclination fixed to $i=30$\degr . Overall, residuals are quite homogeneous, however locally they can reach values larger than $\pm50$~km~s$^{-1}$, especially in the very center. This demonstrates that the HCG 31 A + C system cannot be described as a single rotating disk, despite exhibiting a rotational pattern. This reinforces the hypothesis proposed by A07, where they suggest that what we are observing in the central region is the merging of the HCG 31 A and HCG 31 C galaxies.

There are also two sub-structures, A1 and A2, in the central region (Fig.~\ref{map_vel}, left panel). These sub-structures are part of the northern tidal tail of the group \citep{verdes-montenegro+05}. These sub-structures are most likely composed of material stripped from member A due to the interaction, which now are falling back into the galaxy (\citealt{mendesdol+06}, \citealt{amram+07}). The slit in these objects does not show a considerable velocity variation in A1, but there is a low amplitude in A2 ($\sim$40\,km s$^{-1}$). The latter is counter-rotating relative to A+C, which suggests that it is currently falling back (A07). In contrast, A1 does not show a considerable velocity gradient. Probably it is rotating tangentially, and its rounded shape may be evidence of that. In any case, the projected distance to the central merger ($\sim$9\,kpc) is quite small and it does not have enough mass to gravitationally separate from the central merger ( $log(M/M_{\odot}) \approx 6.6)$.

For galaxy B, A07 derived the following kinematic parameters; PA\,=\,-135 $\pm$ 3$^\circ$, $i$ = 60 $\pm$ 5$^{\circ}$, and rotational velocity is $\sim$\,120\,km s$^{-1}$. Our velocity gradient (applying an inclination correction assuming $i$ = 60 $\pm$ 5$^{\circ}$) is consistent with the determinations of A07 and also with the HI map presented by \cite{verdes-montenegro+05}.

We also ran MocKinG to obtain the kinematic parameters of this galaxy. The model results are shown in Fig. 4 (bottom panel), with the kinematic parameters obtained being $PA = -134\pm 1$\degr , and rotation curve parameters compatible with a solid body rotation curve with a slope of $4.3\pm 0.2$~km~s$^{-1}$~arcsec$^{-1}$ for an inclination of 58\degr\ fixed from morphology.\footnote{Inclination cannot be be constrained due to parallel isovelocity contours in the velocity field induced by the solid body rotation curves} Residuals are low, within $\pm 30$~km~s$^{-1}$. However, it is noteworthy that an interesting pattern is observed with negative residuals in the central regions and positive residuals in the outer parts.

The southern tidal tail is the most extended structure in our field-of-view, ($\sim$20,\,kpc from E1 to F3). It is composed of three structures, from north to south: E, H, and F. In the HI map of this system (\citealt{williams+91}, \citealt{verdes-montenegro+05}), this tail extends upon member G (not seen in our FoV), but its optical counterpart is formed by different small-scale structures, of the three main structures mentioned above. The velocity gradient shows a somewhat peculiar kinematic behavior (Fig. \ref{pseudoslits}).

Galaxy E presents a steep velocity gradient with amplitude of $\sim$70\,km s$^{-1}$. According to A07, this structure is falling back to A+C because of its counter-rotation. Within this structure we identified two main knots of star formation that we denominated E1 and E2 (Figure~\ref{map_vel}, left panel). The difference in velocity between these two members is about $\sim$40\,km s$^{-1}$, but with a smooth velocity transition between them, suggesting that they are part of the same kinematic entity.

Object H consists of a chain of three star-forming knots aligned with the tail. This faint structure has the highest velocity in the tail, with a peak of 4080\,km s$^{-1}$ in H3 and amplitude of $\sim$\,40\,km s$^{-1}$ from H1 to H3. The northern part of the gradient seems to be connected to the southern part of the member E gradient. However, our H$\alpha$ map has a very low SNR in the gap between these objects and this connection cannot be strictly verified with our data. As for the southern part of the gradient, this object shows a rapid velocity decrement in a very small projected distance, immediately before the H3 center. The velocity is about 4080\,km s$^{-1}$ at 40 arcsec ($\sim$\,11.2\,kpc) and 4040\,km s$^{-1}$ at 37 arcsec ($\sim$\,10.4\,kpc), that is a 40 km s$^{-1}$ variation over about $\sim$850 pc. This pronounced variation could be explained with the HI velocity map of this tidal tail (\citealt{verdes-montenegro+05}, figure 16) where it is seen that the center of H3 (e5 on \citealt{verdes-montenegro+05}) lies at the edge of a region kinematically detached from the tail.

The object F is composed of three different knots: F1, F2, and F3. This object is the most plausible TDG candidate in HCG\,31 (\citealt{iglesias-paramo&vilchez97}, \citealt{Richer+03}, \citealt{lopez-sanchez+04}, \citealt{mendesdol+06}, \citealt{amram+07}, \citealt{verdes-montenegro+05}, \citealt{gallagher+10}), along with member R \citep{mendesdol+06} (out of our FoV). Member F does not show a velocity gradient in our position-velocity diagram; our velocity gradient seems flat, with no substantial variations in velocity. We speculate that this kinematic behavior is a result to rotation on the plane of the sky. However, this hypothesis has been refuted by A07, arguing that this explanation is very unlikely because of the extended morphology of object F. The velocity of F1 and F2 is $\sim$3980\,km s$^{-1}$, which is consistent with the HI motions, and more like the velocity of the E member. Considering the velocity of member H, the values obtained for member F could lead to misinterpretations about the nature of the tail, which shows the importance of studying 2D velocity fields. Using long-slit data, \cite{lopez-sanchez+04} obtained a velocity gradient for the tidal tail, but they interpreted it as being two different kinematic structures contained in the same object: a short and warped optical tail that ends in H, and an HI tidal tail that connects A+C with G.  

F3 has a velocity of $\sim$4030\,km s$^{-1}$. A smooth gradient between the south of F2 and F3 is seen on our velocity gradient, suggesting that they are part of the same object. Comparing our velocity field with the HI velocity map of the tail we see that F1, F2, and F3 are part of the region kinematically detached from the tail. This explanation favors the scenario in which F is a TDG in formation.

\subsection{Velocity dispersion}

The H$\alpha$ velocity dispersion ($\sigma_{int}$) (Fig.~\ref{map_vel}, right panel) was obtained by using the parameters derived from a single gaussian fitting performed by {\sc ifscube} and a deconvolution of instrumental and thermal dispersion.

Objects A1, A2, B, E, H, and F show low values of velocity dispersion (5\,km s$^{-1}$ < $\sigma_{int} <$ 30\,km s$^{-1}$) (Fig.~\ref{map_vel}, right panel). However, high-resolution data is required to understand the internal kinematics of these sources in detail, given that the MUSE resolution (50\,km s$^{-1}$) is not high enough to search for expanding shells using diagnostic diagrams; therefore, our determinations on these values are upper limits for these sources. Nonetheless, according to \cite{moiseev+15}, these velocity dispersion values (10\ km $s^{-1} \ < \sigma_{int}$ < 30\ km s$^{-1}$ ) are quite typical for \hii\ galaxies and giant \hii\ regions (\citealt{Firpo+11}).

The A+C complex is the most interesting zone in the velocity dispersion map. It hosts the highest values of $\sigma_{int}$ in the HCG\,31 system, with a peak of $\sim$95\,km s$^{-1}$, and its distribution is not spatially correlated with H$\alpha$ contours.  The two main H$\alpha$ knots in the central region show similar velocity dispersion values of $\sim$50\,km s$^{-1}$.

It should be noted that the complex A+C shows double-peaked H$\alpha$ profiles (A07), which cannot be resolved with the spectral resolution of our data.

\subsection{Extinction: A 2D view of the dust distribution in HCG 31}


The values of E(B-V) range from 0 to $\sim$ 0.4. In some spaxels we detected negative value, which indicates that the coefficient \ha/\hb \ is below the theoretical value of 2.86. However, this value implies specific conditions ($n_e = 100 \ cm^{-2}$, $T_{e} = 10^{4} K$) which is not exactly the case for every zone in HCG\,31 (\citealt{lopez-sanchez+04}). Thus, with different conditions, different values of the coefficient \ha/\hb \ were determined. In these cases, we assumed that the extinction was null. The highest E(B-V) values were in an H$\alpha$ knot that is part of member A. The mean values of E(B-V) for each member of HCG\,31 are presented in Table \ref{table_physpar} col (7) for the map average and col (8) for the integrated spectra. Our values are consistent, within the uncertainties, with previous determinations using spectroscopic data (\citealt{lopez-sanchez+04}) or from estimates performed with the HI density column (\citealt{williams+91}). However, our current analysis significantly improves the spatial coverage of this system.

Objects A1, A2, and A3 show very similar values of E(B-V), considering their uncertainties. In those regions our values of the SNR in H$\alpha$ and H$\beta$ are relatively low ($\sim$ 5-10), and many spaxels with negative extinction are found, limiting the number of spaxels with confident determinations of E(B-V). Therefore, in these regions we measured the extinction near the most intense H$\alpha$ knot, maximizing the SNR and minimizing the uncertainties.

Objects A and A+C lie in the region with the highest SNR on our FoV, because of the overlapping of two MUSE fields (fields 1 and 2 on Figure~\ref{fovs}). Thus, these objects present the lowest uncertainties in our E(B-V) determinations.

\begin{figure*} 
\centering 
\includegraphics[width=0.47\textwidth]{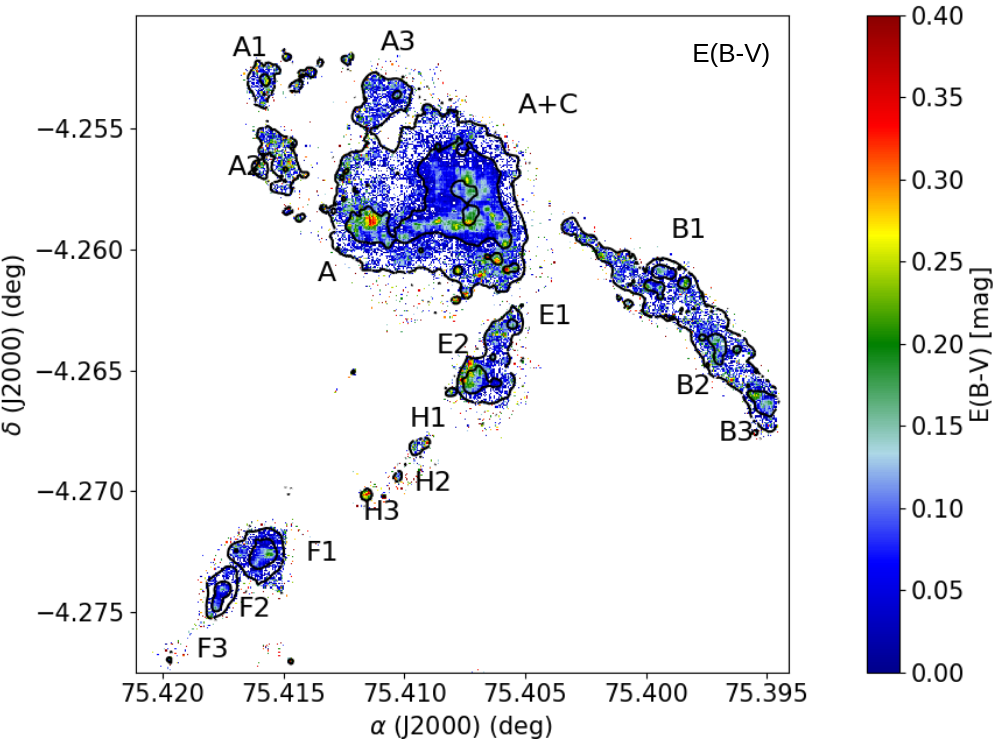}  
\includegraphics[width=0.47\textwidth]{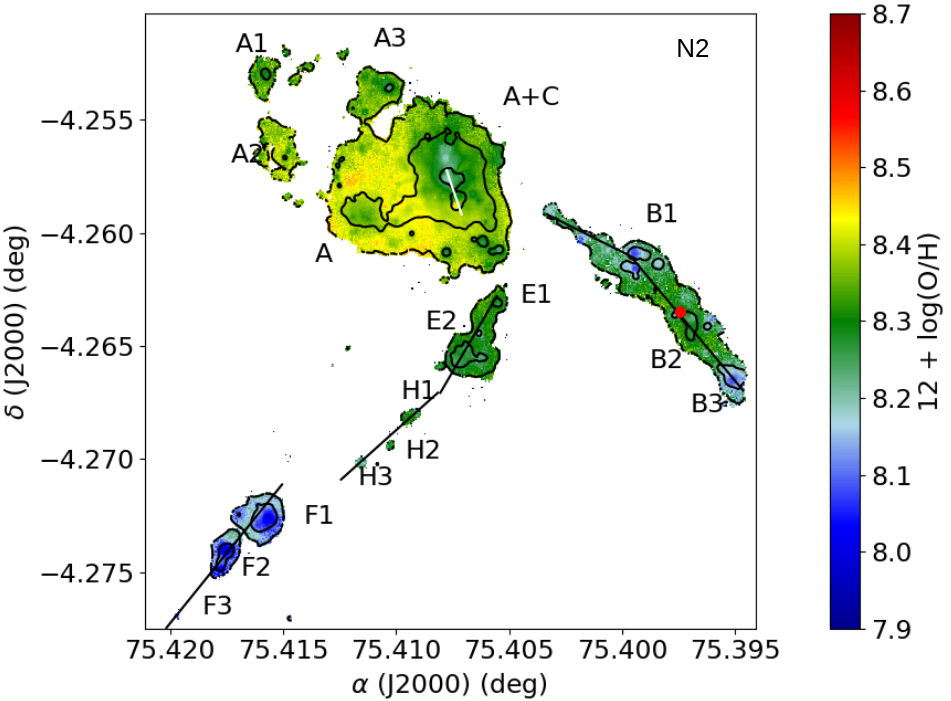}   
\includegraphics[width=0.47\textwidth]{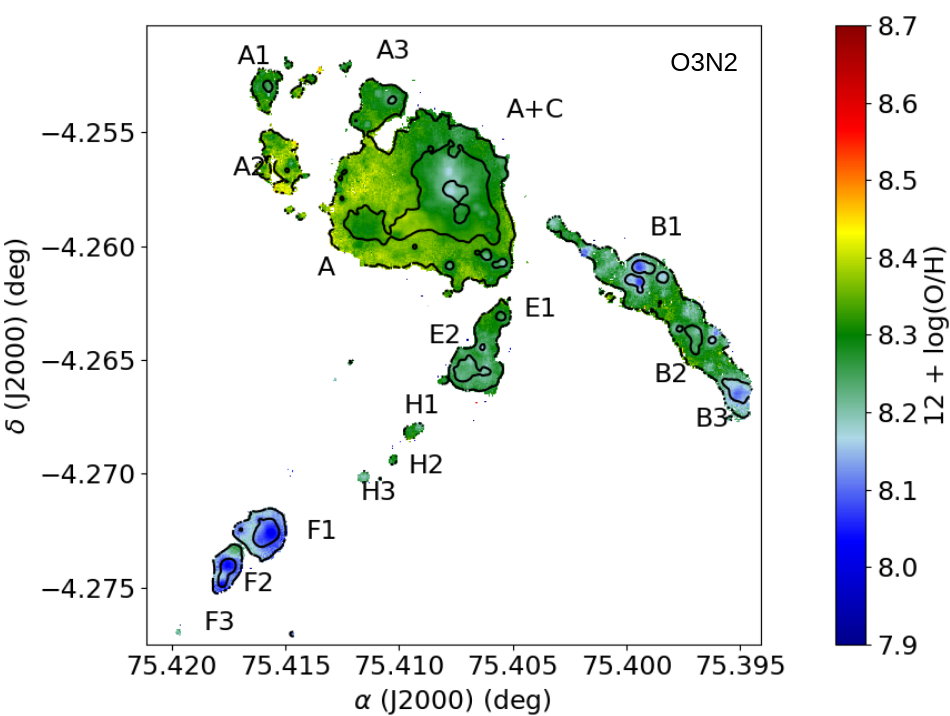}   
\includegraphics[width=0.47\textwidth]{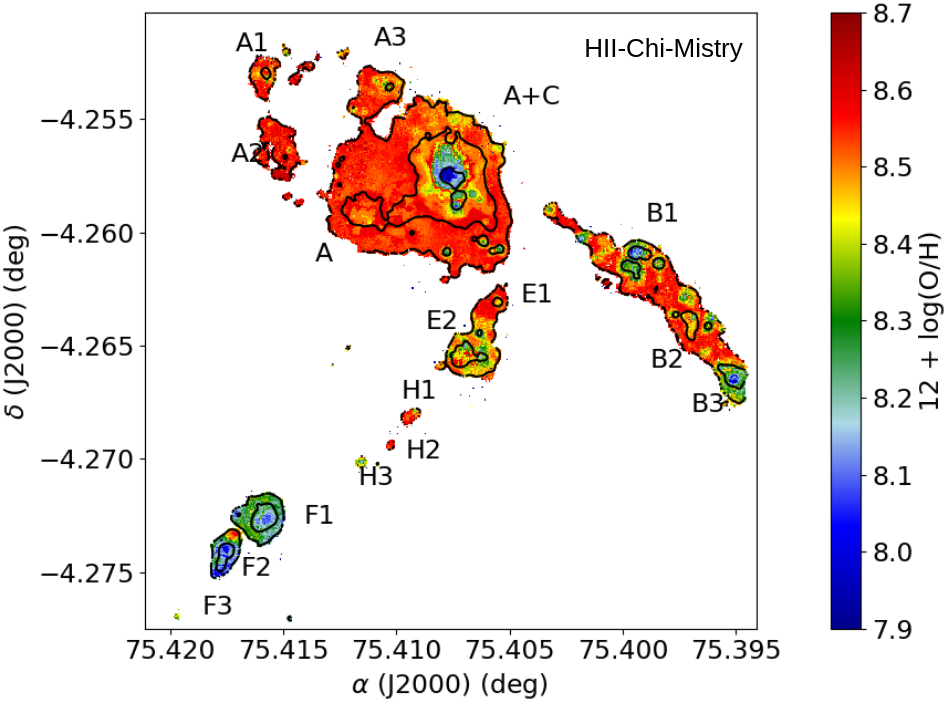}  
\caption{Top left panel: E(B-V) map of HCG\,31. White contours represent H$\alpha$ in emission. The group shows low values of E(B-V) except in the central zone, where E(B-V) shows a peak in a knot of galaxy HCG\, 31 A.
Top right panel: Oxygen abundance map obtained with the strong-line empirical calibrator N2 black lines represent the pseudo slits used to derive the oxygen abundance gradient, the red circle represent the center used for the gradient of the galaxy B. Bottom left panel: Oxygen Abundance map obtained with the strong-line empirical calibrator O3N2.
Bottom right panel: Oxygen abundance map obtained with the code \hcm.} 
\label{ext_abundances} 
\vspace{0.5cm}
\end{figure*}

Object A, in particular, hosts the knot with the highest extinction on HCG\,31, with a value of E(B-V) $\sim0.40 \pm 0.06$. For A+C the absorption is lower, and it is spatially correlated with the H$\alpha$ emission. In the south of A+C, there are several knots with high extinction (E(B-V) $\sim0.4$) that are clearly distinguished in Figure~\ref{ext_abundances} (top left panel). 



\begin{table*}
\begin{tabular}{llllllll}
\multicolumn{1}{c}{Object} & \multicolumn{2}{c}{12 + log(O/H)}                                       & \multicolumn{3}{c}{log Mass [$M_{\odot}$]}                                                  & \multicolumn{2}{c}{E(B-V)}                                             \\ \hline
                   (1)        & \multicolumn{1}{c}{N2 + O3N2 (2)} & \multicolumn{1}{c}{\hcm      (3)} & \multicolumn{1}{c}{FADO (4)} & \multicolumn{1}{c}{LS04 (5)} & \multicolumn{1}{c}{M06 (6)} & \multicolumn{1}{c}{Map average (7)} & \multicolumn{1}{c}{Integrated spectra (8)} \\ \hline
A1                         & 8.34 $\pm$ 0.22               & 8.52 $\pm$ 0.05                         & 6.94                     & ...                      & 8.63                    & 0.19 $\pm$ 0.12                 & 0.17 $\pm$ 0.09                        \\
A2                         & 8.39 $\pm$ 0.21               & 8.55 $\pm$ 0.04                         & 8.05                     & ...                      & ...                     & 0.20 $\pm$ 0.16                 & 0.16 $\pm$ 0.10                        \\
A3                         & 8.30 $\pm$ 0.21               & 8.52 $\pm$ 0.04                         & 7.18                     & ...                      & ...                     & 0.18 $\pm$ 0.17                 & 0.18 $\pm$ 0.13                        \\
A                          & 8.35 $\pm$ 0.21               & 8.55 $\pm$ 0.02                         & 8.73                     & ...                      & ...                     & 0.16 $\pm$ 0.08                 & 0.11 $\pm$ 0.05                        \\
A+C                        & 8.25 $\pm$ 0.21               & 8.31 $\pm$ 0.02                         & 8.61                     & 9.05                     & 9.49                    & 0.12 $\pm$ 0.02                 & 0.13 $\pm$ 0.01                        \\
B                          & 8.31 $\pm$ 0.21               & 8.45 $\pm$ 0.02                         & 8.65                     & 8.90                     & 9.28                    & 0.15 $\pm$ 0.08                 & 0.10 $\pm$ 0.06                        \\
E                          & 8.23 $\pm$ 0.21               & 8.42 $\pm$ 0.04                         & 7.37                     & 7.67                     & 8.03                    & 0.19 $\pm$ 0.10                 & 0.20 $\pm$ 0.10                        \\
F1                         & 8.06 $\pm$ 0.21               & 8.17 $\pm$ 0.04                         & 6.85                     & ...                      & 7.86                    & 0.10 $\pm$ 0.04                 & 0.08 $\pm$ 0.02                        \\
F2                         & 8.10 $\pm$ 0.21               & 8.12 $\pm$ 0.09                         & 5.97                     & ...                      & 7.35                    & 0.07 $\pm$ 0.06                 & 0.06 $\pm$ 0.03                        \\
F(F1 + F2)                 & 8.08 $\pm$ 0.21               & 8.15 $\pm$ 0.07                         & 6.90                     & 7.59                     & 7.98                    & 0.09 $\pm$ 0.05                 & 0.07 $\pm$ 0.03                       
\end{tabular}
\caption{Physical parameters for the members of HCG 31. (Col. 1) ID of the object (Col. 2) Oxygen abundance obtained by the average of N2 and O3N2 methods. (Col. 3) Oxygen abundance obtained by \hcm. (Col. 4) Oxygen abundance obtained by \hcm. (Col. 4) Mass determination performed by FADO (Col. 5)  Mass determined by \protect\cite{Lopez-sanchez+10}. (Col. 6) Adopted mass obtained from the K band luminosities taken from \protect\cite{mendesdol+06}. (Col.7) E(B-V) obtained from an average over the map and (Col. 8) E(B-V) obtained from the integrated spectra of each galaxy. }
\label{table_physpar}
\end{table*}

In object B, the extinctions of its three main knots seem similar, with B2 being the most extinguished. In this galaxy we use the same procedure as in A1, A2 and A3 to obtain the extinction, we averaged the E(B-V) values for the points belonging to the peak in H$\alpha$. It should be noted that this method only gives us an approximation of the extinction. The bridge at the NE of this object shows a mean extinction value of E(B-V) $\sim0.12 \pm 0.08$. The two main knots in object E show similar values of E(B-V), with E2 being the one with the highest extinction and where the H$\alpha$ emission is strongest. For the main body of E, the extinction was E(B-V) $\sim0-0.1$. In member H the emission of H$\alpha$ and H$\beta$ was very weak and had a low SNR. Thus, the uncertainties determining E(B-V) are very high in its three main knots. In object F, the extinction seems to be slightly lower than in the rest of the knots in HCG\,31. In F1 and F2 there is a clear spatial correlation between H$\alpha$ emission and extinction.

\subsection{Electron Density}

\begin{figure}
    \centering
    \includegraphics[scale=0.28]{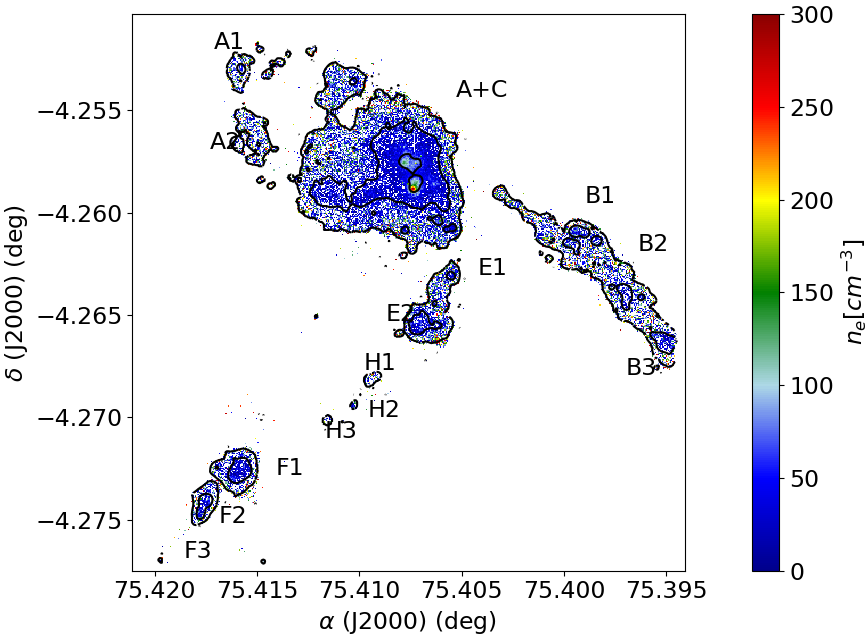}
    \caption{Map of electron density derived with the [$S_{II}$] 6717,6731 $\AA$ lines. The contours are $H\alpha$ in emission. Almost the whole group shows low densities with $n_{e} < 100 [cm^{-3}]$. In central knots of the A+C it is seen a peak in the density with a knot showing $n_{e} \sim 250 [cm^{-3}]$.}
    \label{densitymap}
\vspace{0.3cm}
\end{figure}

In Figure~\ref{densitymap} we show an electron density map, derived from the line ratio of the sulfur doublet \sii $\lambda \lambda  \ 6717, \ 6731 \ \AA$, and the equations presented in \cite{Perez-montero14}. The black contours represent H$\alpha$ emission.

In general, electron densities in HCG\,31 are in the low-density region (n$_{e}$<100 \ cm$^{-3}$), with a peak in the knots located in members A and C. The knot in member A has an average electron density of n$_{e}$ $\sim$ 170 $\pm$ 30 cm$^{-3}$ with a peak of $\sim$ 300 cm$^{-3}$, and the knot in C has an average electron density of 100 $\pm$ 70 cm$^{-3}$ with a peak of $\sim$ 140 cm$^{-3}$. In all the other members, the electron density remained below 100 cm$^{-3}$. These measurements were consistent with the values known for extragalactic \hii \ regions (n$_{e}$ < 500 \ cm$^{-3}$ \citealt{Bresolin+05}).

Several authors have studied the effects of interaction on electron density. For example, \cite{Krabbe+14} studied seven pairs of interacting galaxies and determined that these systems showed higher electron densities (n$_{e}$ = 24-532 cm$^{-3}$) than  isolated galaxies (n$_{e}$ = 40 - 137 cm$^{-3}$). Our values are consistent with those ranges.

\subsection{Oxygen abundance determinations}

To have a complete view of the oxygen abundance in HCG\,31, we obtained oxygen abundance maps using the strong-line empirical calibrators, N2 and O3N2. The abundance maps obtained with these calibrators are presented on the top right and bottom left panels of Figure~\ref{ext_abundances}, respectively. In addition, in the bottom right panel of Figure \ref{ext_abundances} we show, the corresponding oxygen abundance map derived by code {\sc HII-CHII-Mistry}.

In Figure~\ref{ext_abundances} (top right panel) the slits used to obtain the metallicity gradients of the system are shown (see section \ref{oxygen_grad_results}).
In this figure we observed a slight transition from \met\ $=  8.23 \pm 0.16$ (N2) in HCG\,31 C to \met $=  8.40 \pm 0.16$ (N2) in HCG\,31 A. Inspecting Figure~\ref{ext_abundances} (bottom left panel), especially in the central region of HCG\,31, we observed a very similar trend, with a more metallic knot in the south (knot A). 

Using GMOS-IFU data, \cite{Torres-flores+15} studied the central zone of HCG\,31, and found a metallicity gradient in the line that connects the main burst of star formation in HCG\,31 A and C. Our maps suggest the same behavior for oxygen abundance, especially for N2 and O3N2. 

Objects A1, A2, and A3 have very similar values of \met \ spanning a range of 8.3 $<$ \met $<$ 8.4. Taking into account the uncertainties, we consider that these three objects have the same metallicity, with values similar to HCG\,31 A (we note that this trend was observed for all calibrators). This finding is consistent with the tidal origin of these sources, as proposed by several authors (\citealt{amram+07}, \citealt{lopez-sanchez+04}) 

Galaxy B showed a fairly homogeneous distribution of metallicity at the surroundings of the knot B2 with \met $\sim 8.3 $  and a drop to lower metallicities at B1 and B3 with \met \ $\sim 8.1$. The bridge that connects B and A+C seemed to have the same metallicity as the main body of B. 

The mean values of the oxygen abundance for sources E and H are \met \ $8.25 \pm 0.16$ and \met \ $8.20 \pm 0.18$, respectively. These values are similar to the abundances of the central complex A+C, suggesting a tidal origin for sources E and H, similar to objects A1, A2, and A3.

The member F is the most metal-poor object, while F1 and F2 presented the same metallicity for all calibrators. The mean value of metallicity was \met $\sim8.06$.

The highest metallicities were detected for the main body of galaxy A and for the region between the knots of galaxy B, with values of \met $\sim 8.55$ (Fig. \ref{ext_abundances}, bottom right panel). This value is considerably higher than the values of these regions obtained with the N2 and O3N2 calibrators, where the metallicities span a range of 8.2 $<$ \met $<$ 8.4. This difference could be due to the dependence with respect to the ionization parameter or some relative abundances, such as the nitrogen-to-oxygen ratio.

\subsection{Oxygen abundance gradients}
\label{oxygen_grad_results}

We proposed to study the metal distribution of HCG\,31 in two structures: galaxy B and the southern tidal tail. In order to carry out this analysis, we simulated two different slits passing through these structures (Fig. \ref{ext_abundances}, top right panel). In the southern tidal tail, we used the same slit as defined in section \ref{radvel_results}. In galaxy B we split the slit into two parts to cover the connection between the knots B1, B2 and B3, including the bridge between B and A+C. The slits had a width of 1 arcsec and we binned the gradients in  1 arcsecs to match the seeing (seeing $\sim$1 arcsecs).

\begin{figure*} 
\centering 
\includegraphics[width=0.8\textwidth]{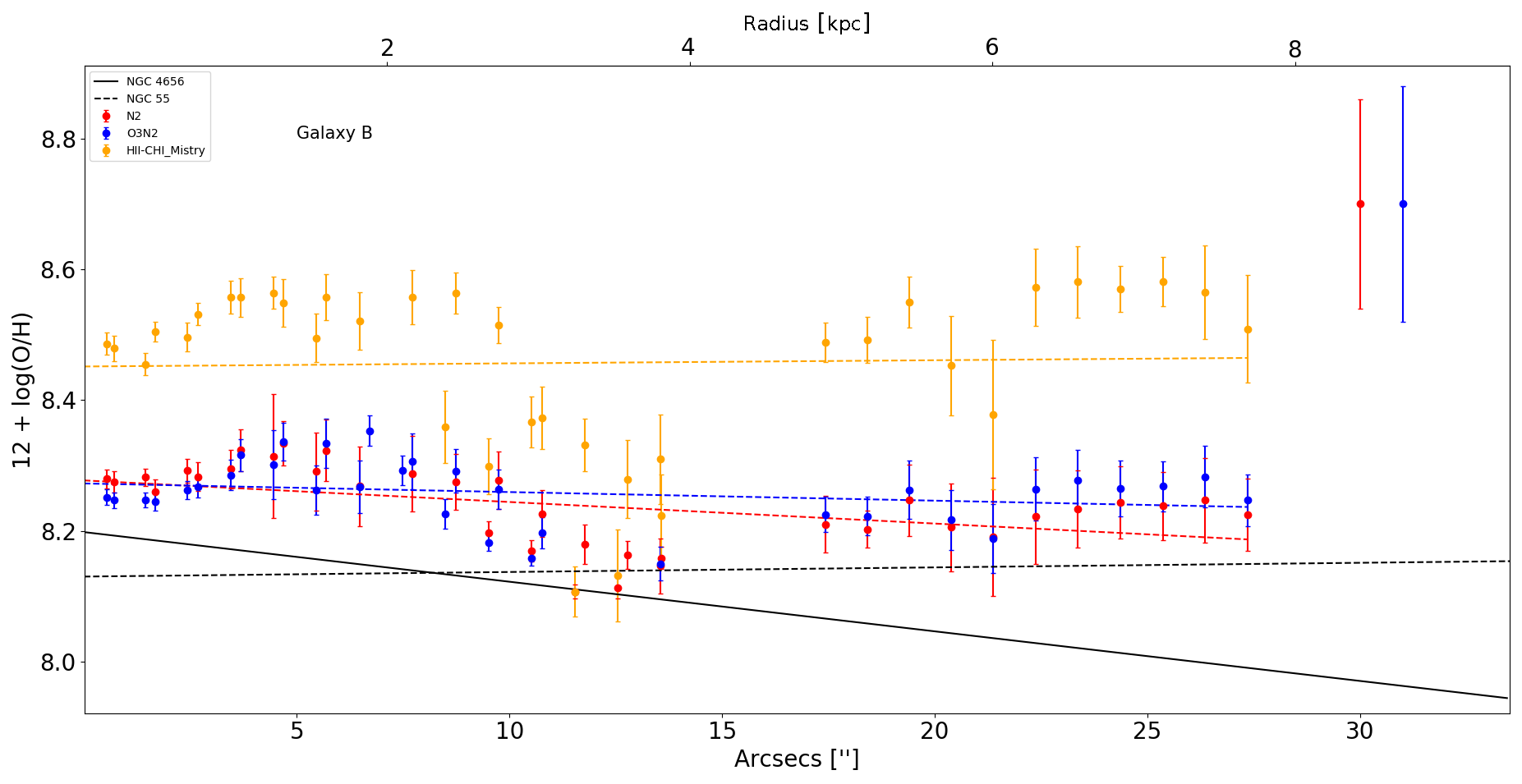}   
\includegraphics[width=0.8\textwidth]{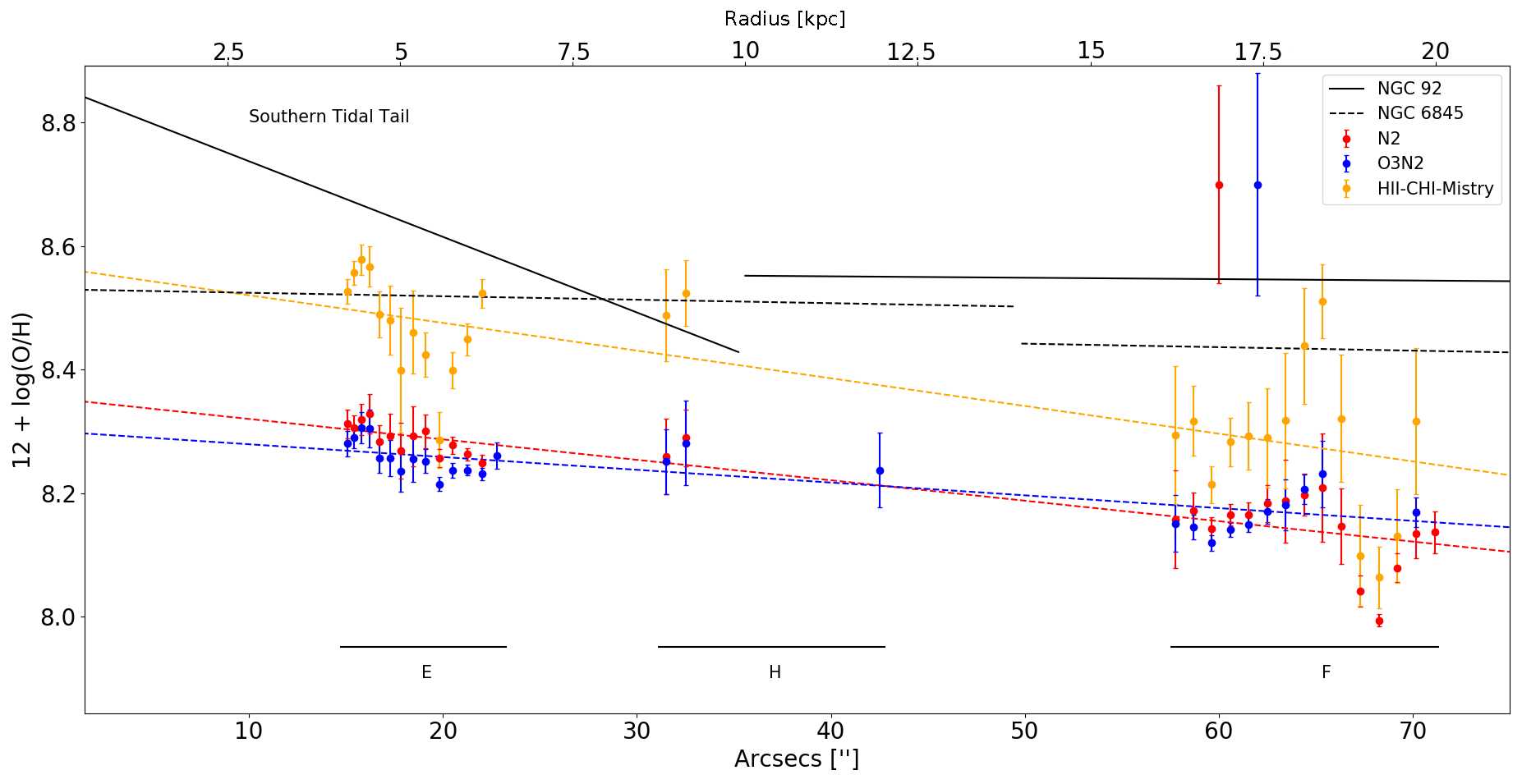} 
\caption{Metallicity gradients and their fit parameters derived for HCG\, 31; (top) Galaxy B and (bottom) southern tidal tail. For comparation we include the gradients of other similar objects in both panels. For more details see section 5.5.} 
\label{met_gradients} 
\vspace{0.5cm}
\hspace{0cm}
\end{figure*} 

The oxygen abundance gradients derived for the different slits are presented in Figure~\ref{met_gradients}. The data were fitted using the package {\sc curve$\_$fit} of {\sc scipy} \citep{Virtanen+20} which allowed us to consider the uncertainties in performing the linear regression. The parameters obtained for the fits are listed in Table \ref{gradients_parameters}. Error bars in Figure~\ref{met_gradients} represent the propagated error coming from flux uncertainties. At the top right of Figure \ref{met_gradients} we show the dispersion associated with each calibrator (0.16 dex for N2, red bar; 0.18 dex for O3N2, blue bar). The segmented lines represent the linear fit performed. (yellow, blue and green colors represent the fit on the \hcm, O3N2 and N2 calibrators respectively).

Inspecting the gradient of galaxy B, we noted a somewhat steep gradient in the main body of the galaxy , with a peak uncorrelated with the center of the galaxy. This offset was about 5 arcsec (1.4 kpc), and it was also observed in the \hcm \  estimates. 

For the bridge of galaxy B, which starts at around 17 arcsecs on Figure~\ref{met_gradients} (top panel), a quite flat metallicity gradient was observed. 

We performed a linear fit on the metallicity distribution of member B, using the measurements on its main body and bridge and the results are shown in Table \ref{gradients_parameters}. We found a gradient of $\alpha = -0.012 \pm 0.002  [dex/kpc]$, with the N2 method, suggesting a flat oxygen distribution which is expected for a galaxy with interaction signatures (\citealt{Kewley+10}, \citealt{rich+12}). 

We compared this metal distribution with results obtained for galaxies NGC\,4656 and NGC\,55, which displayed similar morphological types (SB(s)m, taken from NED) and whose gradients were studied by  \cite{munoz-elgueta+18} and \cite{magrini+17} respectively (Fig.\ref{met_gradients}, left panel). We found slopes similar to those systems, suggesting flat metal distributions. However, the metallicities in NGC\,5656 and NGC\,55 were lower than HCG\,31 B by $0.1-0.2$ dex.

The drop in the central metallicities observed in HCG\,31 B is an interesting feature that could be related to gas inflows in the galaxy. \cite{sanchez+14} found similar results for some galaxies of their sample. They proposed that in many cases this drop was produced by the presence of a star formation ring in the nuclear region of the galaxy. This scenario is difficult to prove in HCG\,31 B because its position is nearly edge-on, therefore we cannot discard it. However, we can speculate that a gas inflow induced by the interaction with the HCG\,31 A+C complex is diluting the central metallicities and triggering star formation at the central zone of this galaxy, producing the central drop in the metallicity gradient.

In the bottom panel of figure \ref{met_gradients} we show the oxygen abundance gradient of the southern tidal tail, which includes members E, H and F, and on the top right are represented the associated dispersion of the empirical calibrators (0.16 dex for N2 and 0.18 dex for O3N2). The gradient begins at the midpoint between knots A and C (HCG 31
A+C complex) because some authors suggested that this tail was
formed from material detached from galaxy C \citep{amram+07}.
On Figure \ref{met_gradients} (bottom panel) we include the oxygen abundance distributions of other tidal tails located in compact group galaxies, namely NGC 92 \citep{torres-flores+14} and NGC 6845 \citep{olave-rojas+15}. We note that NGC 92 and NGC 6845 are spiral
galaxies that belong to groups at a less advanced interaction stage
(no strong evidence of merger yet in these systems). The gradient reveals that the tidal tail of HCG 31 is less metallic than
NGC 92 and NGC6845 by $\sim$0.4 dex and $\sim$0.3 dex, respectively and that the scale lengths are quite different, the tidal tail of NGC 92 starts at a radius of $\sim$10 kpc and extends for $\sim$25 kpc; for NGC 6845 the tail starts at 14 kpc and extends for $\sim$70 kpc, while the southern tail of HCG 31 starts at $\sim$3 kpc and extends about $\sim$20 kpc.

Flat oxygen abundance gradients can be explained by gas flows
induced during galaxy-galaxy interactions (e.g \citealt{rupke+10} ).



\begin{table}
    \centering
    \begin{tabular}{l|c|c}
    \hline
Method         & Zero point {[}dex{]} & Slope {[}dex/kpc{]} \\ \hline
\multicolumn{3}{c}{Galaxy B}                                \\ \hline
N2             & 8.26 $\pm$ 0.02      & -0.012 $\pm$ 0.002  \\
O3N2           & 8.27 $\pm$ 0.01      & -0.004 $\pm$ 0.001  \\
HII-CHI-Mistry & 8.43 $\pm$ 0.03      & 0.002 $\pm$ 0.001   \\ \hline
\multicolumn{3}{c}{Southern Tail}                           \\ \hline
N2             & 8.35 $\pm$ 0.02      & -0.012 $\pm$ 0.001  \\
O3N2           & 8.29 $\pm$ 0.01      & -0.007 $\pm$ 0.001  \\
HII-CHI-Mistry & 8.61 $\pm$ 0.03      & -0.020 $\pm$ 0.001  \\
\end{tabular}
    \caption{Results of the linear regression for the abundance gradients in galaxy B and the southern tidal tail with the different methods}
    \label{gradients_parameters}
\end{table}

Assuming that was the cause, one would expect to find a velocity gradient through the tidal tail; but, as explained in Section \ref{radvel_results}, the position-velocity diagram through the southern tidal tail revealed a velocity gradient only through objects E and H, while F showed a flat velocity gradient because it was kinematically detached from the tail. These findings suggested that gas flows could be the explanation of the flat gradient only for objects E and H. For member F, a metal-poor gas accretion that triggers star formation \citep{amram+07} could be the explanation for its slightly lower metallicity.

\subsection{H alpha luminosities and Star Formation Rates: Witnessing the stellar birth in a merging system}

Using the H$\alpha$ map, corrected for extinction and distance, we derived the SFR map of HCG\,31, which is shown in the top left panel of Figure~\ref{sfr_mapas}. This map was obtained by using the calibration proposed by \cite{Kennicutt&evans12}.
In the central zone of the system, we obtained a total SFR of $\sim2.99 \pm 0.49 M_{\odot} \ yr^{-1}$ (Fig. \ref{sfr_mapas}, black ellipses, top left panel). Assuming a stellar mass of $log M_{stellar} \sim8.5$ \ \citep{Lopez-sanchez+10}, we obtained a value for the specific star formation Rate (sSFR) of $\sim log(sSFR) \sim -8.22$ for the central zone of HCG\,31. If we plot HCG\,31 in the sSFR vs stellar mass plane (\citealt{Atek+14}) we find that the system lies at the locus of galaxies located with redshifts of $0.7 < z < 1.0$. This fact is indicative of the strong episode of starburst that the system is experiencing, which is similar with respect to the SFRs of galaxies located at higher redshifts. If we consider the $H_{2}$ mass concentrated in the central zone of the system ($2.9 \times 10^{8} M_{\odot}$, \citealt{yun+97}) with the current SFR, 75$\%$ of the $H_{2}$ mass would be depleted in $\sim$100 Myr.

\subsection{H alpha equivalent width and ages}

\begin{figure*} 
\centering 
\includegraphics[width=0.45\textwidth]{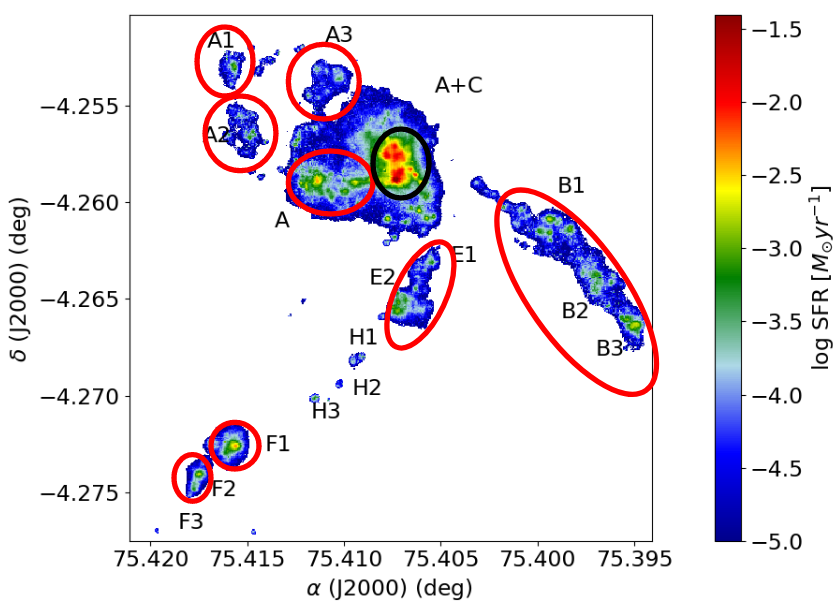}  
\includegraphics[width=0.45\textwidth]{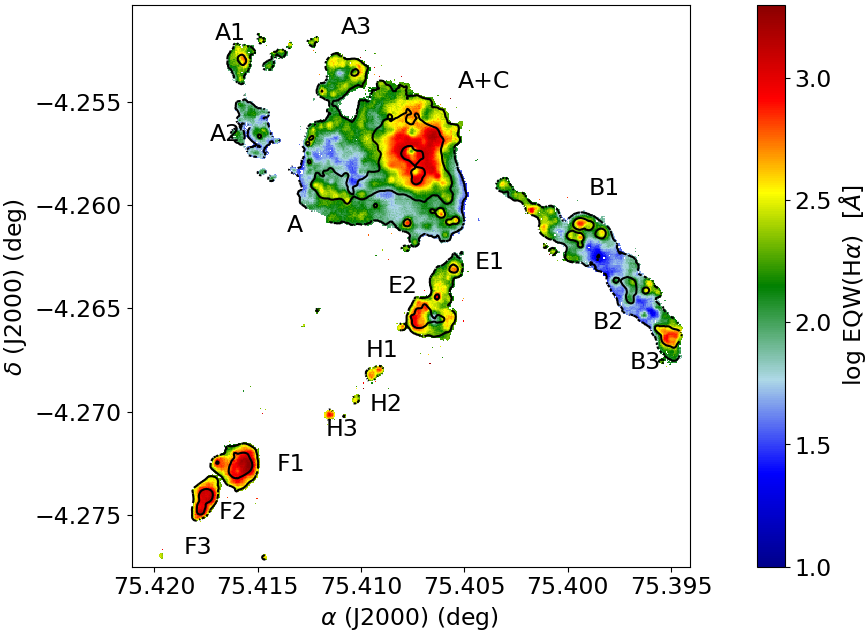}   
\includegraphics[width=0.45\textwidth]{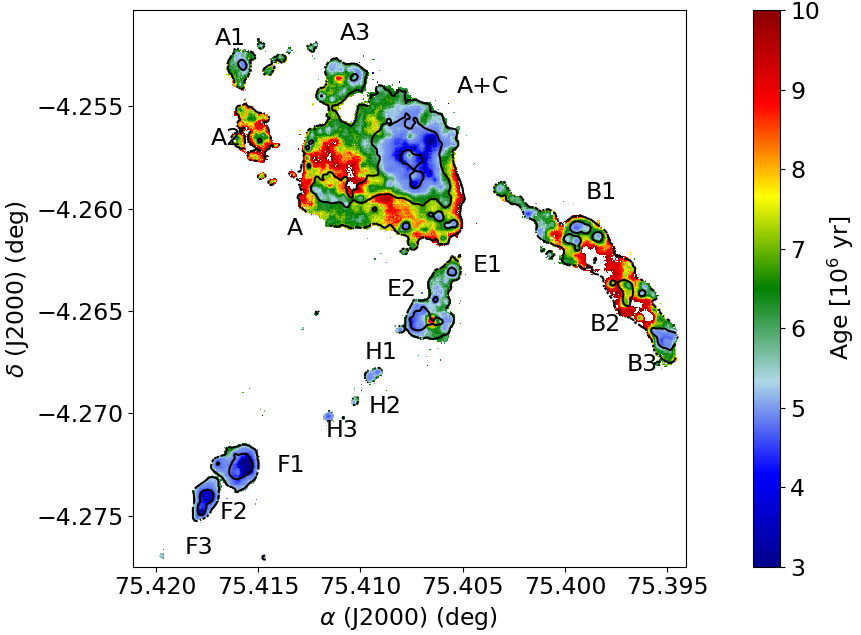}   
\includegraphics[width=0.45\textwidth]{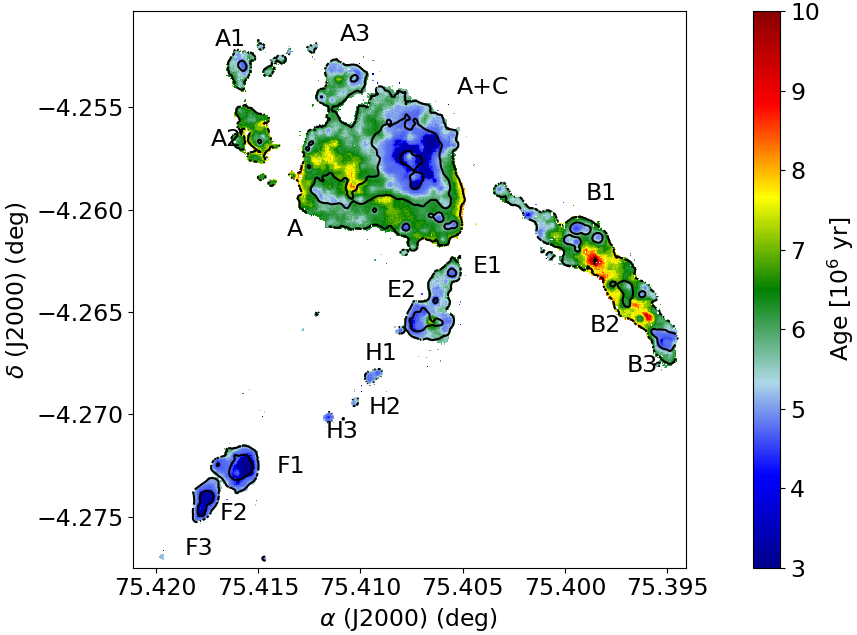}  
\caption{Top left panel: Map of star formation rate for HCG\,31 derived using the calibration of \protect\cite{Kennicutt&evans12}, the ellipses represent the regions where the spectra were integrated (see text). Top right panel: Map of H$\alpha$ equivalent width derived by {\sc FADO}; black contours represent H$\alpha$ in emission. Bottom left panel: Map of ages obtained by interpolating with the models of Starburst99 assuming an instantaneous star formation event. A Salpeter IMF was assumed with mass limits 1-100 $M_{\odot}$ and Z = 0.004. Bottom right panel: Same as bottom left panel but with Z = 0.008.} 

\label{sfr_mapas} 
\vspace{0.5cm}
\end{figure*}

In the top right panel of Figure~\ref{sfr_mapas} we show the EW($H\alpha$) map of HCG\,31, derived from {\sc FADO}. Black contours represent the H$\alpha$ emission. The whole group spans a range of $10 \AA < EW(H\alpha) < 2000 \AA$, where the highest values are associated with the galaxy F and the central merger. On the other hand, the lowest values are associated with member B. The EW($H\alpha$) map is also used to estimate the ages of the star forming knots, by interpolating it with the models of {\sc STARBURST99}. Results are shown in the bottom left and bottom right panels of Figure~\ref{sfr_mapas}, respectively. The ages span a range from 3 Myr to 10 Myr. The younger regions are found in galaxy F and in the central zone of the system with ages $\sim3 Myr$. The older regions are found at the center of galaxy B with ages $\sim 10 Myr$ in the maps with both metallicities.

\subsection{The Wolf-Rayet bump}

Several previous studies have detected WR features in the spectra of HCG\,31 (\citealt{kunth&schild86}, \citealt{lopez-sanchez+04}, \citealt{Lopez-sanchez+10}). \cite{kunth&schild86} were the first authors to report WR features in HCG\,31. They detected $He_{II}$ 4686 $\AA$ and $N_{III}$ 4640 $\AA$ features that were related to the emission of nitrogen Wolf-Rayet stars (WN). \cite{Guseva+00} detected the same lines and added $N_{III}$ 4512 $\AA$  and $Si_{III}$ 4565 $\AA$ at the central zone of HCG\,31. The blending of all these lines in the spectra of a galaxy is the so-called blue bump and it is widely used in the literature to estimate the number of WN stars in a galaxy (\citealt{Vacca&conti92}, \citealt{Lopez-sanchez+10}). In the case of HCG\,31, \cite{lopez-sanchez+04} also detected the blue bump in their long-slit spectra.  In our case, the spectral coverage of our data did not allow us to completely detect the blue bump. However, we could detect another feature that is related to the emission of Carbon Wolf-Rayet stars (WC), the so-called red bump. This red bump is produced by the $C_{IV}$ 5808 $\AA$ line. In the case of HCG\,31 this feature is very weak and difficult to detect. For example, \cite{Guseva+00} and \cite{lopez-sanchez+04} did not detect this line in their spectra, probably because the WR population of this system was dominated by WN stars (\citealt{Guseva+00}, \citealt{Lopez-sanchez+10}). Nevertheless, in our data we can detect this weak feature at the central zone of this system.

\begin{figure} 
\centering  
\includegraphics[width=0.48\textwidth]{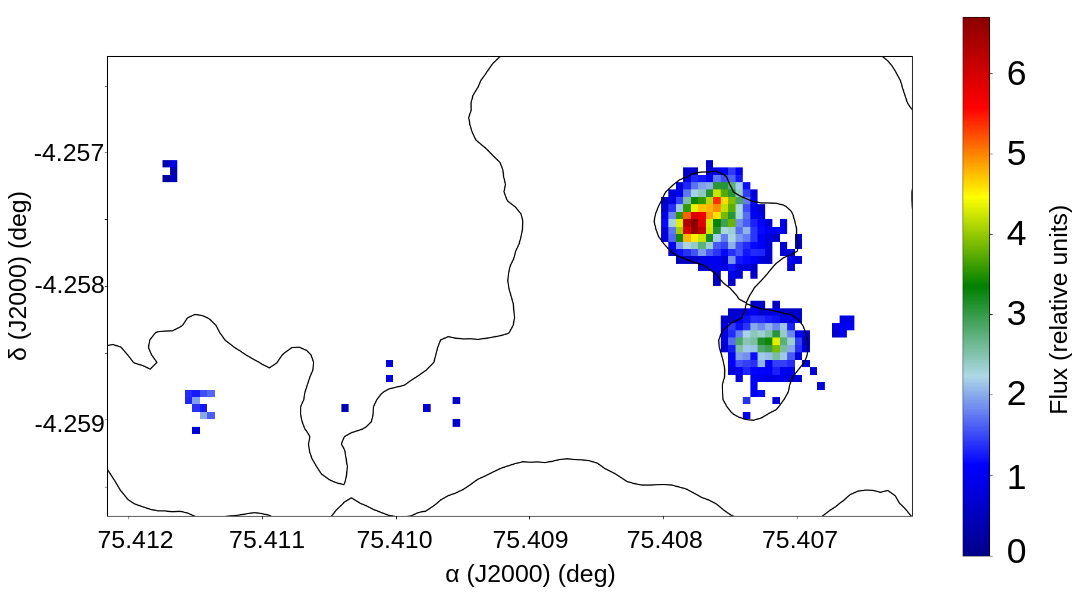} 
\caption{Intensity map of the red bump obtained for the central region of HCG\,31; black contours represent H$\alpha$ in emission.} \label{wr_map} 
\vspace{0.5cm}
\end{figure} 

To map the red bump, we performed a linear fit in the continuum region near to the spectral feature, and then removed it from the data. Using the continuum-subtracted spectra we collapsed the datacube in the range of 5838 $\AA$ to 5934 $\AA$, producing a 2D image. This map is shown in figure~\ref{wr_map}, where we detected the red bump only in the inner central zone of HCG\,31 A+C. It should be noted that the spatial location of the WC stars is consistent with the ages estimated for this zone in the previous section (5-6 Myr).

We integrated all the spectra of the spaxels seen in Figure~\ref{wr_map} and calculated the total luminosity of the red bump. The luminosity obtained is L$_{obs}$(C$_{IV}$ 5808 $\AA$) = (1.09 $\pm$ 0.05) $\times$ 10$^{39}$ [erg s$^{-1}$]. In order to obtain the number of WC stars we used the equations presented by \cite{Lopez-sanchez+10} and assumed an average metallicity of $12 + log(O/H) = 8.40 \pm 0.19$. We obtained a total number of carbon Wolf-Rayet stars (WCE using the notation of \citealt{Lopez-sanchez+10}) of  $N_{WCE} = 492 \pm 75$. This number is twice as high as in previous studies. \cite{Guseva+00} obtained $N(WCE) = 246 $ and \cite{Lopez-sanchez+10} obtained $N(WCE) = 206$. It is important to note that previous studies determined the number of WCE stars using long-slit spectroscopy instead of integral field spectroscopy; therefore, aperture effects cannot be neglected. 

\subsection{Ionization mechanism}

\begin{figure*} 
\centering 
\includegraphics[width=0.75\textwidth]{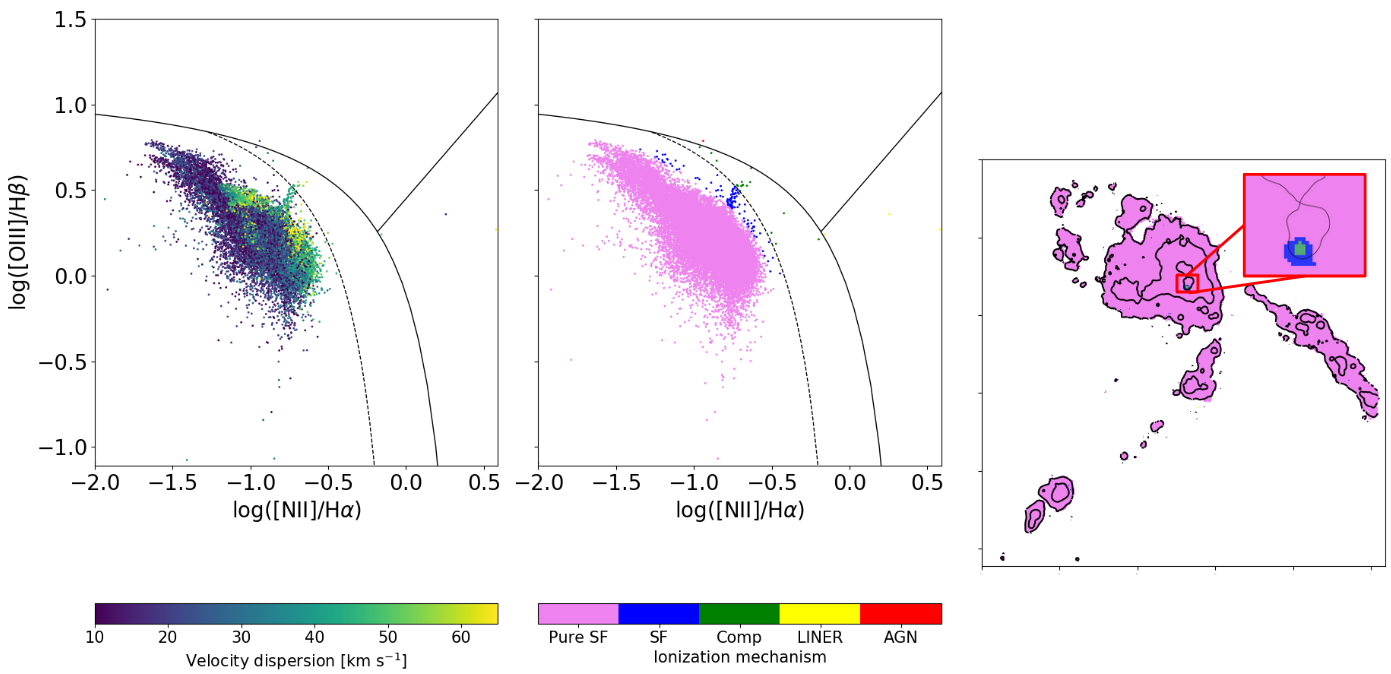}  
\includegraphics[width=0.75\textwidth]{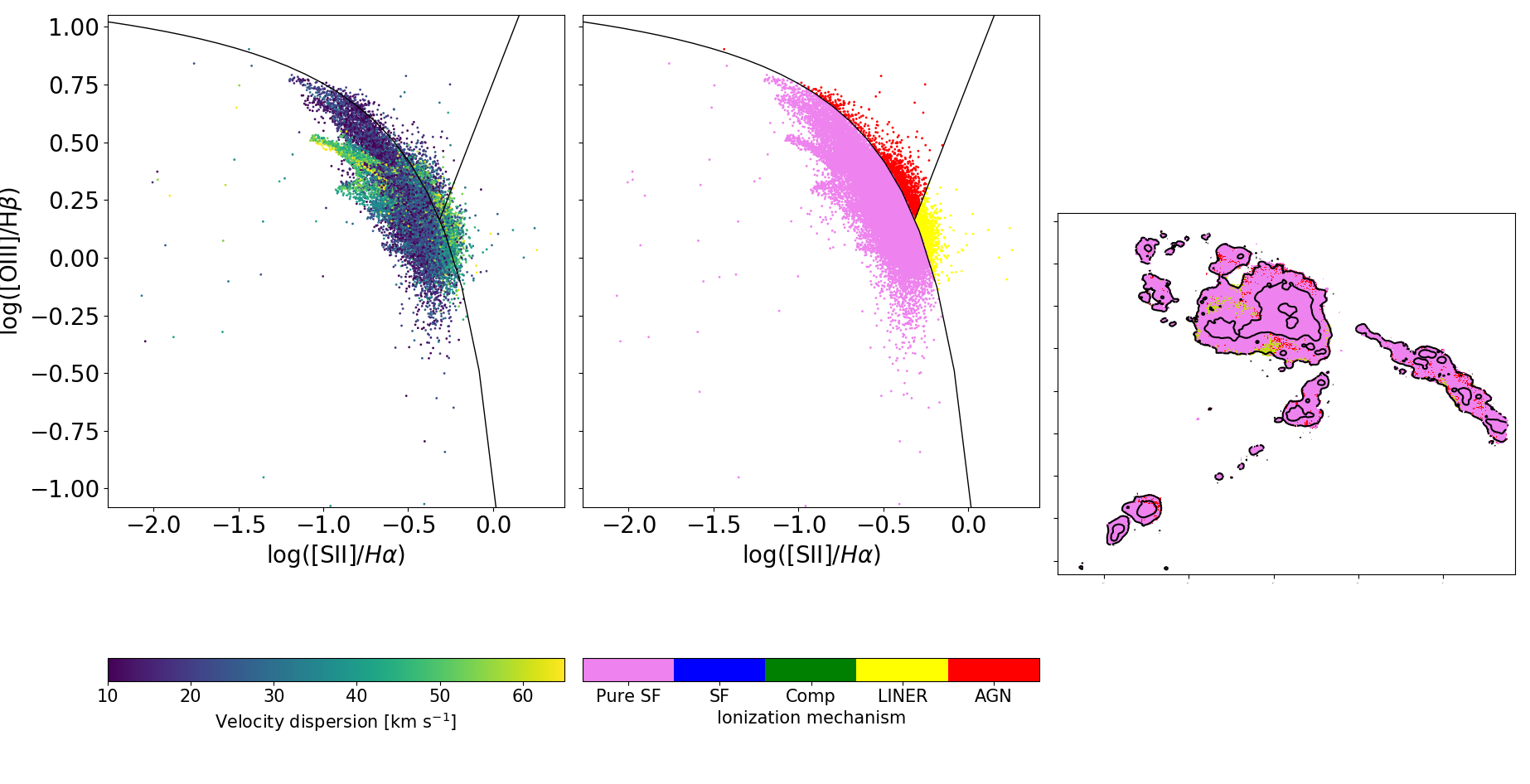}  
\includegraphics[width=0.75\textwidth]{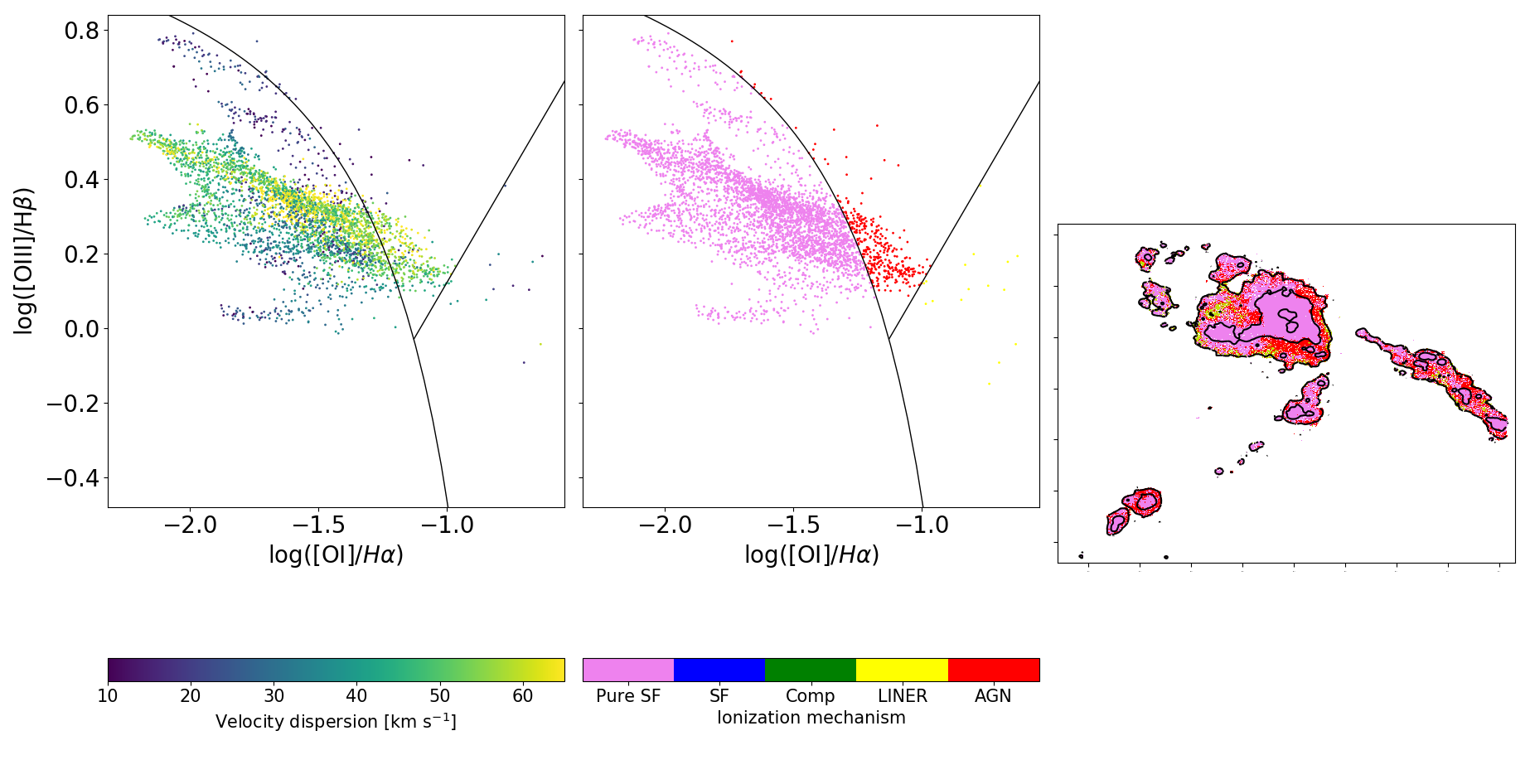}
\caption{BPT diagrams for the whole system. The contours represent $H\alpha$ in emission. Most of the points lie in the SF sequence, while nearly all of the points in the AGN/LINER zones lie in the outskirts of the galaxy. The reason for this may be the low S/N in these regions, or to an eventual contribution from shocks. In left panel we present the diagrams color coded with the velocity dispersion. Middle panel represents the BPT diagrams and right panel the 2D distribution of the points. The borderlines in the BPTs corresponds to the delimitations obtained in \protect\cite{Kewley+01} and \protect\cite{Kauffmann+03} (separation between SF and AGN) and in \protect\cite{Schawinski+07} (separation between AGN and LINER).} \label{BPTs} 
\vspace{0.5cm}
\end{figure*}

We used standard diagnostic diagrams to elucidate the ionization mechanisms in HCG\,31. In Figure~\ref{BPTs} we show the spatially resolved diagnostic diagrams, where the top, middle and bottom panels are the $[O_{III}] / H\beta$ vs $[N_{II}]/ H\alpha$, $[O_{III}] / H\beta$ vs $[S_{II}]/ H\alpha$ and $[O_{III}] / H\beta$ vs $[O_{I}]/ H\alpha$, respectively. We found that most of the system was primarily ionized by massive stars due to recent star formation (pink points in Figure~\ref{BPTs}). Yellow points represent ionization by shocks/LINER, which are negligible, as can be seen in the top panel of Figure~\ref{BPTs}. Red points are associated with AGN as the main ionization mechanism. This scenario is unlikely, because the spatial distribution of the red points is not consistent with the presence of an AGN. If an AGN was the main ionization mechanism, we should find a more central distribution, not a random distribution in the outskirts of galaxies Figure~\ref{BPTs}.

Inspecting the central zone, we found that knot A was in the composite region in the diagnostic ionization diagram that used the $[N_{II}]/H\alpha$ ratio. This outcome could be associated with a violent star formation event that triggered massive stellar winds and shocks. If we inspect the velocity dispersion map ($\sigma_{int}$) of this zone we find that it is $\sigma_{int}\sim 60 km \ s^{-1}$, which is lower than the velocity expected for shocks ($\sigma_{int}>90 km \ s^{-1}$, \citealt{Rich+15}), but higher than the velocity expected for a regular \hii\ region ($\sigma_{int}$ $\sim$ 30 km s$^{-1}$, \citealt{moiseev+15}). However, different regions can be observed along line of sight. Therefore these values of sigma could indicate shocks. Also, as mentioned in previous sections, the widths that we measured were likely to be overestimated due to multiple unresolved components, thus the velocity dispersion that we present is an upper limit for this region. This allowed us to conclude that this knot was ionized by a combination of shocks and star formation. This scenario was supported by a high EW(H$\alpha$), a high H$\alpha$ luminosity (SF evidence) and the line ratios obtained (shocks/composite ionization evidence). The main ionization mechanism for the remaining galaxies is the star formation.   

We also included a color-coded panel for sigma in the three different BPT diagrams. We did not observe a strong correlation between the increase in velocity dispersion and the contribution of shocks, at least in the sulfur and oxygen diagrams (middle and bottom panels, respectively). In the BPT diagram using nitrogen, we did observe a weak correlation between the increase in velocity dispersion and proximity to the composite zone. This may indicate stronger evidence that shocks are contributing significantly to the ionization of the central region of the system.


\subsection{The mass-metallicity relation of HCG 31}
\label{mzr}

It is well established that there is a correlation between the stellar mass of galaxies and their gas-phase metallicities (\citealt{Tremonti+04}, \citealt{Yates+20}). This correlation, known as mass-metallicity relation (MZR), has been widely used to study the formation and evolution of galaxies.
\cite{Weilbacher+03} showed that TDGs did not follow the MZR, thus, plotting this relation for the members of HCG\,31 and inspecting the positions of the different objects should give us clues about their formation. Several works studied the luminosity-metallicity relation of HCG\,31 for the purpose of discriminating among the primordial and tidal dwarf galaxies in this group. \cite{Richer+03} and \cite{lopez-sanchez+04} studied the \mbox{M$_{B}$ vs 12 + log(O/H)} relation for this group and \cite{mendesdol+06} studied the \mbox{M$_{K}$ vs 12 + log(O/H)} relation. In the case of the \mbox{M$_{B}$ vs 12 + log(O/H)} relation, the data suggest that the members, A, B, C and G, are very luminous for their respective metallicity. This could be explained by the fact that the magnitude on the B-band filter is probably contaminated by the luminosity from the star formation processes in the system \citep{lopez-sanchez+04}. In contrast, \cite{mendesdol+06} posited that the \mbox{M$_{K}$ vs 12 + log(O/H)} relation suggested that members, F1, F2, E2, A1, H and R, are TDGs or tidal debris.

To plot the MZR for HCG\,31, we integrated the spectra of nine different objects: A1, A2, A3, A, A+C, B, E, F1 and F2. We ran {\sc FADO} on each spectrum, allowing us to obtain the stellar mass of each galaxy. The metallicities for each spectrum were based on the average of N2 and O3N2, and on \hcm. The results are listed in columns (2) and (3) of Table \ref{table_physpar}. The masses obtained are listed in column (4) of Table \ref{table_physpar}. For comparison, we included other mass estimations derived in previous studies. In column 5 we show the masses obtained by \cite{Lopez-sanchez+10} for the different members of HCG\,31. In column 6 we list stellar masses derived from the $K-band$ magnitudes published by \cite{mendesdol+06}. In this case we assumed a mass-to-light ratio for each member, based on their $g'-r'$ colors (see \citealt{Bell+03}), and assuming a solar $M_{K} = 5.08 \ [mag]$ \cite{Willmer18}.

\begin{figure*}
    \centering
    \includegraphics[scale=0.45]{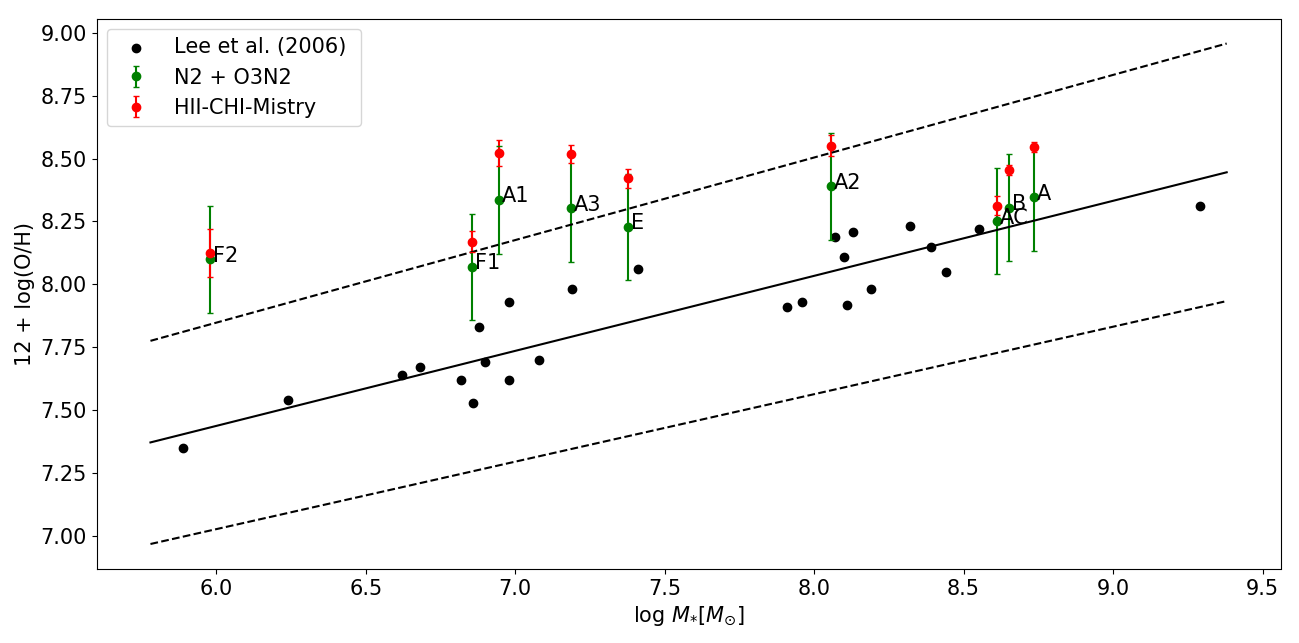}
    \caption{The MZR relation derived for HCG\,31 (green and red points). For comparison, we included the MZR of nearby dwarf galaxies derived by \protect\cite{Lee+06} as black points. The masses considered were calculated by {\sc FADO}. The adopted metallicity was obtained from an average of N2 and O3N2 and the red points stands for \hcm \ determinations.}  \label{MZRs}
\vspace{1cm}
\end{figure*}

As a control sample we used the MZR derived by \cite{Lee+06}, which estimated masses and metallicities for 27 nearby dwarf galaxies. This relation is shown by black points in Figure~\ref{MZRs}. It should be noted that the metallicities considered on this control samples were obtained through the direct method \citep{Lee+03}. In Figure~\ref{MZRs}, we present the MZR of HCG\,31. We see that galaxies A, B and the A+C complex follow the main MZR, which is expected for these kinds of galaxies \citep{mendesdol+06} and also favours the fact that the LZR with the M$_{B}$ band is contaminated by the luminosity of newly formed stars. Other objects seems to be out of the main trend, which is not surprising considering the scenario on which all these objects were formed in the tidal tails, after the first gravitational encounter in HCG 31. (\citealt{amram+04}, \citealt{amram+07}).

\section{Discussion and conclusions}
\label{discussion}

In this paper we present an analysis of the physical and kinematic properties of the complex compact group of galaxies called HCG\,31. Considering the large field of view covered on the current analysis, this study improves our understanding of this system, which has been previously studied by different authors from different observational approaches (\citealt{iglesias-paramo&vilchez97}, \citealt{Johnson+99}, \citealt{gallagher+10}, \citealt{rubin+90}, \citealt{lopez-sanchez+04}, \citealt{mendesdol+06}, \citealt{amram+07}, \citealt{alfaro-cuello+15}). In this section, we discuss the main findings of this work.

\subsection{Abundance gradients and the influence of the environment}
\label{grad_disc}

The metallicity distribution in spiral galaxies has been extensively studied in recent years (\citealt{vanzee+98} \citealt{Bresolin+12}, \citealt{sanchez+14}). Most giant galaxies show a clear abundance gradient, with the center of the galaxy being more metallic than the outskirts. However, several observational studies have proven that interacting galaxies show flatter abundance gradients than isolated galaxies (\citealt{Kewley+10}, \citealt{torres-flores+14}, \citealt{olave-rojas+15}). Numerical simulations showed that this flattening in the gradients was mainly produced by gas inflows towards the nuclear zones that were triggered by gravitational encounters and mergers \citep{Torrey+12}. In addition, observational studies have shown that dwarf galaxies have small abundance gradients or none. (\citealt{roy+96}, \citealt{Hunter&Hoffman99}, \citealt{Croxall+09}, \citealt{Izotov+06}). \cite{pilyugin+15} found that irregular dwarf galaxies could present an abundance gradient depending on their surface brightness profile, breaking the spiral vs dwarf dichotomy, which says that spiral galaxies display strong abundance gradients, whereas irregular dwarf galaxies do not. Recently \cite{James+20} used MUSE observations of the dwarf galaxy, JKB 18, and discovered that it had an inhomogeneous chemical distribution, providing more evidence that not all irregular dwarf galaxies were chemically homogeneous.

In this paper, we present detailed metallicity maps of the HCG 31 system, based on several strong-line empirical calibrators. The excellent spatial coverage provided by MUSE allowed us to derive the metallicity gradient for galaxy B, the southern tidal tail and for the central zone of the system. In the case of galaxy B, we detected an almost flat metallicity distribution with a slope of $\alpha$ = -0.012 $\pm$ 0.002 dex \ kpc$^{-1}$ (N2 calibrator), which provided information about the chemical homogeneity of this system, with no significant variations in its metallicity. This finding is expected for a merging galaxy \citep{rupke+10a}.

An interesting feature in the gradient of this galaxy was the central drop in metallicity seen at $\sim$5 arcsec. This kind of a decrease seems to be similar to those detected by \cite{sanchez+14} in several galaxies in their sample. However, those authors found no correlation between these decreases and interaction features in the galaxies. Here we propose three different scenarios to explain this central drop in HCG 31 B.

\cite{amram+07} found that the receding and approaching sides of the rotation curve of this galaxy did not match. The most prominent disagreement in the rotation curve was at the inner 5 arcsec. \cite{amram+07} argued that this inner disagreement might be the sign of a bar, which could be inducing radial motions of gas that were flattening the central metallicities. However, it should be noted that there was no correlation between the presence of a bar and the central drop in metallicity \citep{sanchez+14}. Another possibility is that the galaxy is currently accreting metal-poor gas, which is inducing the starburst and producing the central drop in metallicity. This scenario is unlikely because a galaxy that is accreting material shows a drop in metallicity of $\sim$-0.5 dex \citep{sanchez-almeida+14}, and the central drop that we detected was $<$-0.2 dex regardless of the method used. Accretion of a partially enriched gas of the group could explain this drop instead of pristine gas.

Lastly, one other possibility is that the galaxy hosts a central star-forming ring which constitutes evidence of radial gas flows induced by resonance processes. This is the scenario that \cite{sanchez+14} used to explain the central drop in the galaxies of their sample. However, they did not find evidence of signs of interaction or bars in these galaxies. In our case, galaxy B showed both signs, thus, we speculated that a combination of the presence of a bar and gas accretion could be responsible for the central metallicity drop in this galaxy.

A nonphysical explanation could be that the drop is artificial and just a random effect of the strong line methods used. Indeed, if we consider the 0.16 dex uncertainties of the strong-line methods the drop dissapears. We used three different methods to plot the gradient (N2, O3N2 and \hcm ) and with each method we observed the same effect, which supports our conclusion that the observed drop is real. A determination of the metallicity gradient of galaxy B with the direct method is required to verify the presence of this central drop.

For the southern tidal tail we also detected a flat metallicity distribution with a slope of $ \beta = -0.012 \pm 0.001$ dex \ kpc$^{-1}$  (N2 calibrator) and $\beta = -0.007 \pm 0.001$ (O3N2 calibrator) dex \ kpc$^{-1}$. Member F seems to be slightly less metallic than members E and H, which could be a consequence of the current metal-poor gas accretion. Several works have studied the metallicity gradients of tidal tails (\citealt{Chien+07} \citealt{torres-flores+14}, \citealt{olave-rojas+15}) and their results suggested that tidal tails showed flat metallicity gradients. A similar effect was found for larger radii (R $>$ R$_{25}$) in normal disk galaxies \citep{sanchez+14}. However, in the literature search, we did not find a metallicity gradient determination in a system similar to HCG 31; thus, we can only compare our gradient with gradients of tails in more massive systems. The origin of this flattening effect is generally attributed to streaming motions inside the tail that redistribute the gas. If this is the case, one would expect to find a velocity gradient along the tail, and indeed, this is exactly what we found for this tail. An increasing velocity was observed from object E to H; but, in object F we did not detect the same gradient. We observed a flat velocity gradient for this galaxy, which indicated counter-rotation with respect to the tail (\citealt {verdes-montenegro+05}, \citealt{amram+07}).


\subsection{Are we witnessing ionization induced by shocks in HCG 31?}

\begin{figure*}
    \centering
    \includegraphics[scale=0.62]{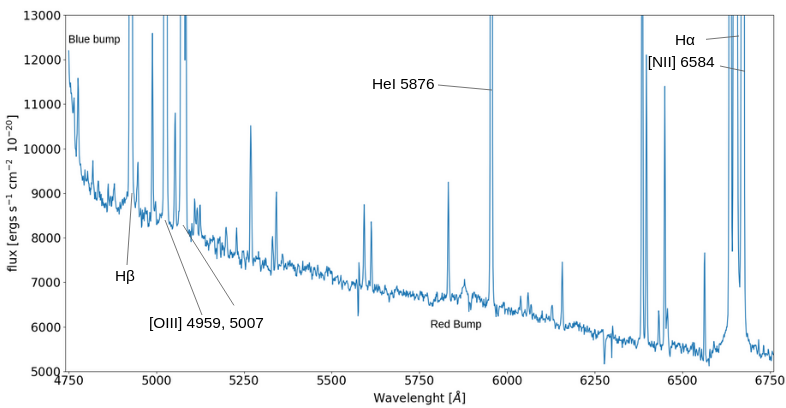}
    \caption{Fine structure of the integrated spectrum in a 1$"$x1$"$ box centered at knot A. The spectrum shows WR features, a clear red bump and the beginnings of a blue bump. The most intense emission lines are labeled.}
    \label{specA}
\vspace{0.5cm}
\end{figure*}

An interesting region located in the central A+C complex is the brightest H$\alpha$ knot. This knot is usually referred to as the nucleus of galaxy HCG\,31 A (\citealt{Torres-flores+15}) and it shows several interesting properties. It is the only region in the entire A+C complex that seems to be ionized not only by star formation, it shows the highest electron density and H$\alpha$ emission of the entire group \citep{alfaro-cuello+15} and the highest metallicity of the system.

Normally, one expects to find low metallicities in HII regions having high H$\alpha$ luminosities, because the starburst is likely fuelled by an influx of metal-poor gas \citep{sanchez-almeida+14}. However, we detected a high oxygen abundance for this region of \met \ = 8.22 $\pm$ 0.19, on average. According to \cite{oey&kennicutt1993}, systematic variations in nebular density can lead to differences in metallicities of up to 0.5 dex. This effect cannot be discarded for this knot, because it shows the highest electron density in the entire system with $n_{e}$ $\sim$300 cm$^{-3}$.

This knot is also the region with the highest oxygen abundance of the system, and it is well known that WR stars can contribute to enriching the interstellar medium (\citealt{Perez-Montero+13}, \citealt{Kehrig+13}). To confirm the existence of such stars we integrated the spectrum over a box of 1" $\times$ 1" centered at knot A. The resulting spectrum is presented in Figure~\ref{specA}. A characteristic red bump was observed. \cite{Krabbe+14} proposed that the stellar winds and mass loss of WR stars were the cause of one of the denser and metal-rich HII regions in their sample, a scenario very similar to the case of this knot.

The velocity dispersion on this knot was $\sim$55 km \ s$^{-1}$, which seems to be consistent with shocks ionization(\citealt{Rich+15}). In this context, this velocity could be expected considering the location of this knot in the BPT diagram, which lies in the composite zone.

A good way to disentangle the ionization mechanism is by using 3D diagnostic diagrams \citep{Kewley+19}. Unfortunately, we could not apply this type of diagnostic to our data for two main reasons: one of the axis of the 3D plot needs the galactocentric radius which is difficult to determine in a merging system such as HCG\,31, and more reliable determinations of the velocity dispersion are needed to avoid over estimations because of the multiple components. High-resolution spectroscopic data is needed to rule out the presence of shocks.

\subsection{Unveiling the star formation history of a complex compact group.}

\begin{figure*}
    \centering
    \includegraphics[scale=0.4]{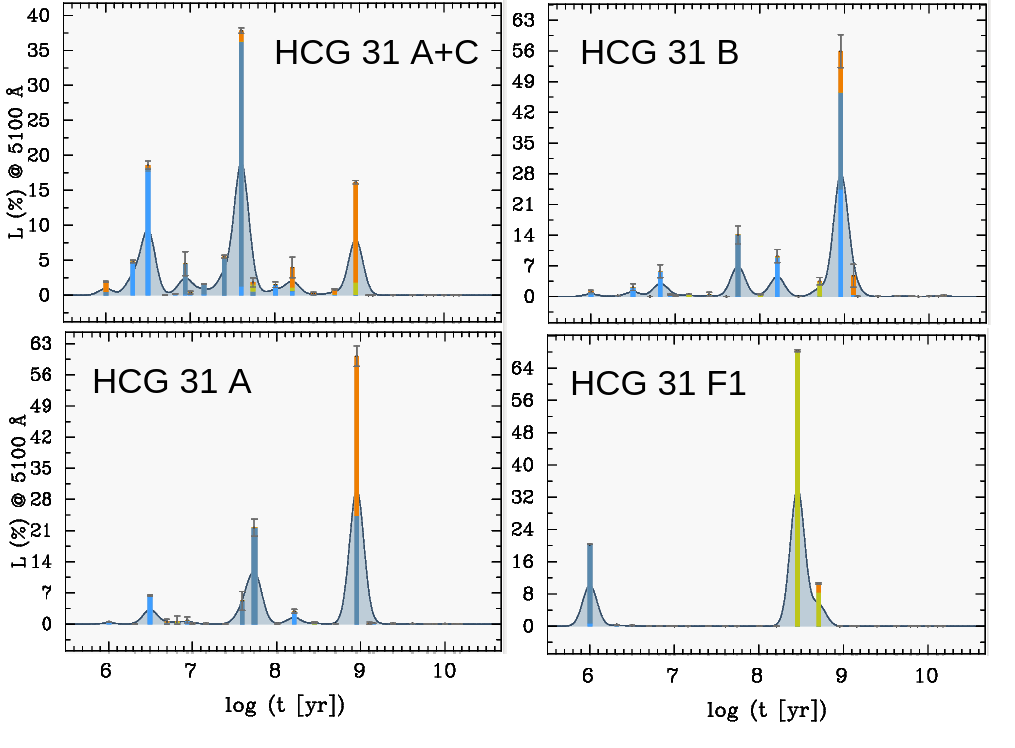}
    \caption{Luminosity fraction at the normalization wavelength (5100 \AA) as a function of age for galaxies HCG\,31 A+C, HCG31 A, HCG\,31 B and HCG\,31 F1. The colors, light-blue, blue, light-green and orange represent metallicities of 0.02 Z$_{\odot}$, 0.2 Z$_{\odot}$, 0.4 Z$_{\odot}$ and 1 Z$_{\odot}$, respectively. The shaded area represent the Akima-smoothed \protect\citep{akima70} version of the populations. This representation provides an illustration of the star formation history of the different galaxies.}
    \label{SFHs}
\vspace{0.5cm}
\end{figure*}

Our new spectroscopic information about HCG\,31 provides evidences that allow us to uncover the star formation history of this interacting system. 

In order to determine the stellar populations in HCG 31 we ran FADO over the integrated spectra of each galaxy. Figure~\ref{SFHs} shows the luminosity fraction at the normalization wavelength (5100 \ \AA) for each of the SSPs fitted, which illustrates the star formation history of each galaxy. We showed SFHs only for the galaxies with the best SNR in the continuum (A+C, A, and B). We also included the SFH of galaxy F1. 

The SFH of the A+C complex revealed that the peak of star formation occurred from $\sim$ 10$^7$ yrs to $10^8$ yrs ago. We find that this age is consistent with the date of the first encounter, as reported by \cite{Johnson+99} (400 Myrs). In addition, the system experienced another starburst event between $\sim$ 10$^{6}$ yrs and 10$^7$ yrs ago, which is completely consistent with the ages of the star forming bursts obtained in this work.

The SFH of galaxy A is very similar to that of the A+C complex, with an important underlying old stellar population at $\sim$ 1 Gyr, another peak at 10$^{7}$ yrs to 10$^{8}$ yrs and the current star formation event at 10$^{6}$ yrs to 10$^{7}$ yrs ago. It should be noted that the SFH of galaxy A represents only the SFH of the extended body that is seen in the optical images. It is not a representation of the SFH of the galaxy HCG\,31 A as whole. It is not possible to completely separate galaxies A and C because they probably overlap each other \citep{amram+07}. For galaxy F1, we found a very interesting aspect in its SFH, with only two peaks of star formation, one at $\sim$ 10$^{8}$ yrs to 10$^{9}$ yrs and the other at about $\sim$ 10$^{6}$ yrs. We did not find evidence of a very old stellar population with ages $>$ 1 Gyr. Our results are in very good agreement with the photometric ages of the SSCs found by \cite{gallagher+10}, who reported no evidence of an old population in this object. 

In summary, all the galaxies of the group display an underlying stellar population of $\sim$1 Gyr old and are currently forming stars. The first encounter between HCG\,31 A and HCG\,31 C occurred about $\sim$ 400 Myrs ago. Galaxy F1 shows a bimodal age distribution, with intermediate and young stellar populations. 

This type of analysis has also been done for other interacting systems. For example, \cite{Buzzo+21} studied the stellar populations of the merging system NGC 1487. They found an age distribution very similar to ours for HCG\,31, with a peak in the younger population which correlates with the current SF episode of the system and an intermediate population with ages of $\sim$ 1-5 Gyrs.

\subsection{Is HCG 31F really a TDG candidate?}

One of the most frequent questions in the literature  about this system concerns the nature of object F. In optical images, it appears as a bright tidal object located in the southern tail of the system. The photometry of this object was quite uncertain because of contamination from a nearby projected star. In addition, \cite{Hunsberger+96} did not consider the F galaxy as a TDG candidate although they did consider the other five objects (E1, E2, A1, A2 and A3) as TDG candidates. It is probably that they did not obtain good photometry on F and discarded it from the analysis. Using H$\alpha$ imaging, \cite{iglesias-paramo+01} proposed that objects F1, F2 and F3 were the most likely TDG candidates in HCG\,31, based on their H$\alpha$ luminosities ($>10^{39}$ ergs \ s$^{-1}$) and their large projected distances from the parent galaxy. A TDG candidate needs to fulfill two main conditions (based on \citealt{Weilbacher+03}): (i) it is quite metallic for its mass (out of the MZR), and (ii) it shows independent kinematics decoupled from the tail. To reach the first condition, many works studied the luminosity-metallicity relation for HCG\,31 using filters $B$ and $K$ (\citealt{Richer+03}, \citealt{lopez-sanchez+04}, \citealt{mendesdol+06}). In all these works, objects F1 and F2 showed high metallicities for their luminosities. In section \ref{mzr} we note that F1 and F2 seemed to be out of the main MZR of dwarf galaxies, and it is well established that object F has a tidal rather than a primordial origin.
The H$\alpha$ kinematics of member F is quite peculiar, because it shows no rotation in any axis \citep{amram+07}, and, our radial velocity map (Section \ref{radvel_results}) confirmed that result. The HI kinematics of F \citep{verdes-montenegro+05} shows that it is kinematically decoupled from the tail, with an axis of rotation that is perpendicular to the axis of the tail, suggesting that this object was already decoupled from the tail. \cite{amram+07} proposed that this object was accreting material from the tail, but that scenario is difficult to prove, and simulations are needed to confirm it. In this work we detect high SFRs and high EW(H$\alpha$) (young ages) in member F, which is fully consistent with this scenario. 
In conclusion, the true nature of object F is not yet fully understood. We confirmed its tidal origin using the MZR relation, and it is very likely that the object is already decoupled from the tail based on its HI kinematics.

\section{Summary}
\label{summary}

On this paper we perform a deep analysis of the Hickson Compact group 31 using IFS data observed with MUSE. We used different maps for analyzing the kinematics, ionization mechanisms, physical properties, chemical abundances, SFRs, ages, among others. Our most remarkable results are:

\begin{itemize}
    \item  The group shows a complex velocity field, with clear evidences of an ongoing merger process in the central region between HCG 31 A and HCG 31 C galaxies.
    
    \item The central zone shows the higher velocity dispersions, with velocities up to $\sim \ 90 \ km \ s^{-1}$. These high velocities are spatially correlated with the merging zone of the system. A more detailed kinematics analysis, that consider the resolved physical properties of the system, is needed to understand the origin of these high velocities. 

    \item The electron density shows a peak in the central zone of the system with $n_{e} \sim 300 cm ^{-3}$. This peak is associated with a very peculiar knot located in the Galaxy HCG\,31 A which probably hosts ionization produced by shocks.
    
    \item  The main ionization mechanism through the whole group is the star formation, with a small contribution of shocks only at the nucleus of galaxy A.

    \item The oxygen abundance is mainly high for the mass of the galaxies, and it shows a flat distribution across the different galaxies. This suggests gas mixing through the whole group probably triggered by the merger.
    
    \item  The star formation rate is high compared to the mass of the system. There are two simultaneous bursts of star formation: In the central zone and Galaxy F.
    
    \item There is a prominent population of carbon Wolf-Rayet stars in the central zone of the group. The presence of these stars is in perfect agreement with the ages obtained from the H$\alpha$ equivalent width. We cannot estimate the population of WN stars.
    
    \item The Mass-Metallicity relation confirms the tidal origin of objects E, H and F.
    
    \item We reconstruct the star formation history of the youngest population in HCG 31. The ages obtained are in perfect agreement with the scenario of a first encounter $\sim$ 400 Myr ago and a current strong episode of star formation.

\end{itemize}

\section{Acknowledgements}

DGE and STF acknowledges the financial support of the Dirección de Investigación of the Universidad de La Serena, through a ‘Concurso de Apoyo a Tesis 2019’. 
We warmly thank Mariane Girard for preliminary analysis on the HCG 31 data in the frame of her master internship.
\section{Data Availability}

The data used on this work is available via the ESO science archive facility \url{http://archive.eso.org/scienceportal/home/.}



\bibliographystyle{mnras}
\bibliography{references} 





\bsp	
\label{lastpage}
\end{document}